\documentclass[a4paper,fleqn,usenatbib,useAMS]{mnras}
\usepackage{graphicx}	
\usepackage{amsmath}	
\usepackage{amssymb}	
\usepackage{multicol}   
\usepackage{bm}		    
\usepackage{pdflscape}	
\usepackage[version=4]{mhchem}
\usepackage[T1]{fontenc}
\usepackage{xcolor}
\usepackage{tikz}
\usepackage{rotating}
\usepackage{lscape}
\usepackage{graphicx,epsfig,epsf,graphics,epstopdf}
\usepackage{hyperref}
\usepackage{mhchem,hhline}
\usepackage[normalem]{ulem}
\usepackage{caption}
\usepackage{float}
\usepackage{amsmath,amstext,amssymb}
\usetikzlibrary{arrows}
\usetikzlibrary{positioning}
\usetikzlibrary{shapes,snakes}
\usepackage[figure,figure*]{hypcap}

\usepackage[T1]{fontenc}
\usepackage{ae,aecompl}
\usepackage{txfonts}

\title[Identification of Swift J1728.9--3613 as black hole]{Swift J1728.9--3613 is a black hole X-ray binary: spectral and timing study using {\it NICER}}
\author[Saha, Mandal \& Pal]{Debasish Saha$^{1}$, Manoj Mandal$^{2}$, Sabyasachi Pal$^{2}$\thanks{E-mail: sabya.pal@gmail.com} \\
$^{1}$Department of Physics, Indian Institute of Science Education and Research Bhopal, Madhya Pradesh, 462066, India \\
$^{2}$Midnapore City College, Kuturia, Bhadutala, Paschim Medinipur, West Bengal, 721129, India \\
}

\begin{document}
\label{firstpage}
\pagerange{\pageref{firstpage}--\pageref{lastpage}}
\date{}
\maketitle

\begin{abstract}
We study different timing and spectral properties of the new Galactic X-ray transient Swift J1728.9--3613 using {\it NICER} and {\it Swift}, discovered by the Burst Alert Telescope (BAT) on the Neil Gehrels {\it Swift} Observatory. The source went through multiple transitions to different spectral states during the outburst, and the complete evolution created a ``$q$''-shaped track in the hardness-intensity diagram. A partial hysteresis is also observed in the RMS-intensity diagram, which is another well-defined phenomenon of black hole transients. In SIMS, power density spectra were dominated by broadband noise components, and two type B QPOs were detected. We have fitted 1--10 keV energy spectra obtained from {\it NICER} observations that were performed during the outburst, and the temporal evolution of spectral parameters were studied. On MJD 58584.69, a small-scale reflare happened, and we observed that the spectral index decreased to a much lower value associated with finite changes in other spectral parameters also, and the 1--10 keV averaged flux also increased. We observed that the innermost radius of the accretion disc was almost constant during the soft state, which corresponds to the Innermost Stable Circular Orbit (ISCO).
We have measured the lower limit of mass of the compact object to be $\sim$4.6 $M_\odot$, considering a non-spinning black hole binary system, by fitting 1--10 keV {\it NICER} spectra with the {\it diskbb} component. The soft-to-hard transition occurred when the bolometric luminosity was 0.01 times the Eddington luminosity. Based on our combined study of the evolution of the timing and spectral properties, we conclude that the new source Swift J1728.9--3613 is a black hole X-ray binary.
\end{abstract}

\begin{keywords}
X-rays: binaries -- accretion, accretion discs -- black hole physics -- X-rays: individual: Swift J1728.9--3613
\end{keywords}


\begingroup
\let\clearpage\relax
\endgroup
\newpage

\section{Introduction}
\label{sec:intro}
X-ray transients, such as black holes, are known to stay in quiescence for most of their lifetime and occasionally go into outbursts. During an outburst, the luminosities of transients increase by several orders of magnitude in all wavelengths \citep{Re06}. X-ray transients in an outburst, evolve through different spectral states with changing luminosity \citep{Re06, Be09}. The source remains in the Hard State (HS) at the early stage of the outburst. As the luminosity increases, the hardness ratio (HR) decreases, which indicates a transition toward the soft state (SS). Different intermediate states like hard intermediate states (HIMS) and soft intermediate states (SIMS) are also observed sometimes during this transition \citep{Ho05, Be11}.
The SS can be observed at a wide range of source luminosity, and HR is at its minimum during this state. A source remains in the SS for a much longer duration compared to that in other states. Towards the end of the SS, HR increases, which indicates the transition towards the harder states and the return back towards the quiescence \citep{Ho05}. Sometimes HIMS and (or) SIMS are also observed for a concise duration in the decay phase of the outburst.
 Energy spectra in the HS are characterized by a power-law component with a typical spectral index ($\Gamma$) varying between 1.6 and 2.1, and a cut-off ranges between 20 keV to 100 keV \citep{Mo09}. Soft photons from the disc are up-scattered due to inverse Comptonisation by the optically thin hot electron cloud which is known as ``Corona'' \citep{Gi10}. Accretion discs are relatively cooler during HS. In SS, energy spectra are dominated by strong blackbody radiation from the innermost region of the accretion disc and a very steep power-law component with a typical $\Gamma\gtrsim 2.5$ \citep{Re06} and the inner disc temperature becomes as high as $\gtrsim$ 1 keV \citep{Sh73, Du11}.

Evolution of the source through different spectral states is studied using the hardness-intensity diagram (HID; \citet{Ho01}) that often makes a ``$q$''-shaped hysteresis track traversed in the anticlockwise direction \citep{Ho05, Du11}. The Hardness-RMS diagram (HRD; \citet{Be05}) and the RMS-intensity diagram (RID; \citet{Mu11}) are also studied to visualize the spectral evolution and understand the nature of compact objects. BHCs exhibit hysteresis during the evolution of an outburst in the RID as well \citep{Mu11}.

The timing and spectral properties of BHXRBs show significant evolution during an outburst. Different types of quasi-periodic oscillations (QPOs) associated with noise components are present in the PDS in different states of an outburst \citep{Ca05, Mo16, In19}. Depending on the positions, significance, RMS amplitudes, and type of noise present in the PDS, these QPOs are classified into three categories. Type C QPOs are quite common to be found in BHXRBs. They are typically observed in the HS and HIMS, where the PDS is dominated by strong band-limited noise (fractional RMS is about 30\%). In the frequency range of a few mHz to 30 Hz, they have sharp and narrow peaks with up to 15\%  RMS amplitude \citep{Wi99, Ca04, Mo11, Mo16}. In the SIMS, the PDS is mainly dominated by weak broadband noise components ($\sim$10\%  fractional RMS). Type B QPOs appear with sharp and narrow peaks, occasionally observed in the frequency range of 4--8 Hz \citep{Ca04, Mo11, Sr13, Mo16, St16}. The fractional RMS in the SS is weaker ($\sim$1\%) compared to that in other states. Type A QPOs are usually very rare to observe but sometimes appear in the SS. They appear with a very weak, flat, and broad peak in the frequency range of about 6--8 Hz with a few per cent RMS amplitude \citep{Ho01}.

The X-ray transient Swift J1728.9--3613 (also known as MAXI J1728--360; \citet{Ne19}) was discovered on January 28, 2019 (MJD 58511.05) with the Burst Alert Telescope (BAT) onboard the Neil Gehrels {\it Swift} Observatory \citep{Ba19}. Following the report of \citet{Ba19} and \citet{Ne19}, \citet{Ke19} determined the source position at RA = $17^h28^m58.64^s$, Dec = --36$^{\circ}$14$\arcmin$37.7$\arcsec$ (J2000) with an error radius of 1.7$\arcsec$ (90\% confidence) with {\it Swift}/XRT. \citet{Ne19} detected the source on  2019 January 26 (two days earlier to its discovery) using the {\it MAXI}/GSC telescope but misidentified the source as a nearby X-ray star EXO 1722--363. Analyzing the time scale of the increase in intensity and the evolution of the hardness ratio, \citet{Ne19} suggested that the new transient might be an accreting pulsar or black hole. \citet{En19} made follow-up observations of the new transient with {\it NICER}. In their observation, the timing and spectral properties of the new source were similar to those found in black holes, which were in the SIMS of the outburst.

In this paper, we studied the detailed timing and spectral properties of the newly discovered X-ray transient Swift J1728.9--3613 using {\it NICER} and {\it Swift} data. Observations and analysis methods are discussed in Section \ref{sec:observations}. The results of different timing and spectral studies are given in Section \ref{sec:result}. We finally made our discussions and conclusions about the nature of the transient on the basis of our results in Section \ref{sec:discussions} and Section \ref{sec:conclusion} respectively.

\section{Observation and data analysis}
\label{sec:observations}

\subsection{{\it NICER}}
\label{subsec:{NICER}}
The primary instrument of the Neutron Star Interior Composition Explorer ({\it NICER}) is the X-ray Timing Instrument (XTI), which is an array of 56 photon detectors that operate in the 0.2--12 keV energy range \citep{Ge16}. Since the discovery of the source Swift J1728.9--3613, {\it NICER} monitored the source regularly. In this study, we used all available archival data during the outburst between MJD 58512.64 and MJD 58657.00, having a total exposure of $\sim$175 ks, to study the evolution of the timing and spectral properties and understand the nature of the source. We used the {\tt NICERL2} task to apply the standard screening and calibrations using the latest calibration files. After performing the barycentre correction using the task {\tt BARYCORR}, we extracted light curves and spectra using {\tt XSELECT v2.4e}. For the study of the power density spectrum (PDS), we extracted light curves of different bin times (1 s, 0.1 s, and 0.01 s). We used the response matrix file (nixtiref20170601v002.rmf) and ancillary response file (nixtiaveonaxis20170601v004.arf) for spectral study. Backgrounds were estimated from each observation using {\tt nibackgen3C50} script \citep{Re21}. Details of data products are shown in Table \ref{tab:parameters} along with the results obtained.

\subsection{{\it Swift}}
\label{subsec:swift}
The Neil Gehrels {\it Swift} Observatory is designed mainly to study Gamma-ray bursts (GRBs). It operates in a wide range of energies -- from optical to soft gamma rays with the help of three onboard instruments on it, which are -- the Burst Alert Telescope (BAT; \citet{Ba05}), Ultraviolet and Optical Telescope (UVOT; \citet{Bu00}) and the X-ray Telescope (XRT; \citep{Po08}). We used the monitoring data of BAT to study the evolution of X-ray flux of Swift J1728.9--3613 in 15--50 keV, which is shown in Fig. \ref{fig:light}. XRT data in photon counting (PC) mode are used to study the energy spectra (0.3--10 keV) at the beginning of the outburst. We performed {\tt XRTPIPELINE} task to produce re-calibrated and cleaned level 2 event files. We choose a circular region of radius 10$\arcsec$ to be the source region and another circle of radius 50$\arcsec$ to be the background region away from the source. The level-2 cleaned files and region files (selected by {\tt DS9}) are used to extract light curves and spectra for the source and background, both by using {\tt XRTPRODUCTS}. We have re-binned the spectrum with at least 25 counts per bin using the task {\tt grppha}.
We have used the {\tt HEASOFT v6.28} software for the reduction of {\it NICER} and {\it Swift} data. Light curves were plotted using {\tt FPLOT} in {\tt FTOOLS}.


\section{Results}
\label{sec:result}

\subsection{Chandra localization}
\label{subsec:chan_location}
{\it Swift}/BAT detected a previously uncatalogued X-ray transient (Swift J1728.9--3613) at RA = $17^h28^m56^s$, Dec = --36$^\circ$14$\arcmin$13$\arcsec$ (J2000) with an uncertainty of 3$\arcmin$ (90\% confidence) on MJD 58511.05 (28 January 2019 UT 01:22:25) \citep{Ba19}. {\it Swift}/XRT performed a 1 ks target of opportunity (ToO) observation in photon counting (PC) mode on MJD 58511.70. Using this XRT data, \citet{Ke19} found the updated position of the new bright source at RA = $17^h28^m58^s$.64, Dec = --36$^\circ$14$\arcmin$37$\arcsec$.7 (J2000) with a 1.7$\arcsec$ error radius (90\% confidence), which was about 0.7$\arcmin$ away from the {\it Swift}/BAT reported position.

Chandra observed the source with the Advanced CCD Imaging Spectrometer spectroscopic array (ACIS, \citep{Ga03}) (Obs. ID--21213) with exposure of $\sim$5.07 ks on MJD 58526.40. We analyzed Chandra data to extract the image of the source and to measure the position of the source precisely. The raw data was reprocessed using the task {\tt chandra\_repro} with {\tt ciao} version 4.12. We used the {\tt ciao} task {\tt wavdetect} to determine the position of the source.
We found the position of Swift J1728.9--3613 was at RA = $17^h28^m58^s.6401$, Dec = --36$^\circ$14$\arcmin$35$\arcsec$.321 (J2000) with 0.49$\arcsec$ error radius (90\% confidence). This position is consistent with the {\it Swift}/XRT measured position, which is only 2.52$\arcsec$ away, whereas the {\it Swift}/BAT reported position is about 38.88$\arcsec$ away.

\citet{Hu19} claimed that the optical counterpart of the source was located at RA = $17^h28^m50^s.32$, Dec = --36$^\circ$12$\arcmin$21$\arcsec$ ($\pm$1$\arcsec$, J2000) after initial observation by the 60 cm BOOTES-3/YA robotic telescope. But, with Chandra's observation, it seemed that it was unlikely to be the optical counterpart as it lay almost 167.76$\arcsec$ away from the Chandra measured position.

From the 2MASS\footnote{\href{https://ned.ipac.caltech.edu/}{https://ned.ipac.caltech.edu/}} catalog, we found that there were two sources within the error circle of Chandra. The nearest one was 2MASS J17285855--3614333, located at RA = $17^h28^m58.552^s$, Dec = --36$^\circ$14$\arcmin$33.34$\arcsec$ (J2000) which was only 2.22$\arcsec$ away from the Chandra measured position. The flux obtained in $K_{s}$ band (1.39$\times$10$^{14}$ Hz) from this point source was $13.964\pm0.077$ mag \citep{Sk06}. Another source near the Chandra error radius was 2MASS J17285822--3614353 which was located at RA = $17^h28^m58.223^s$, Dec = --36$^\circ$14$\arcmin$35.34$\arcsec$ (J2000) and was 5.04$\arcsec$ away from the Chandra measured position. The flux from that point source in $K_{s}$ band (1.39$\times$10$^{14}$ Hz) was >13.797 mag \citep{Sk06}.

\begin{figure}
\centering{
\includegraphics[width=8cm]{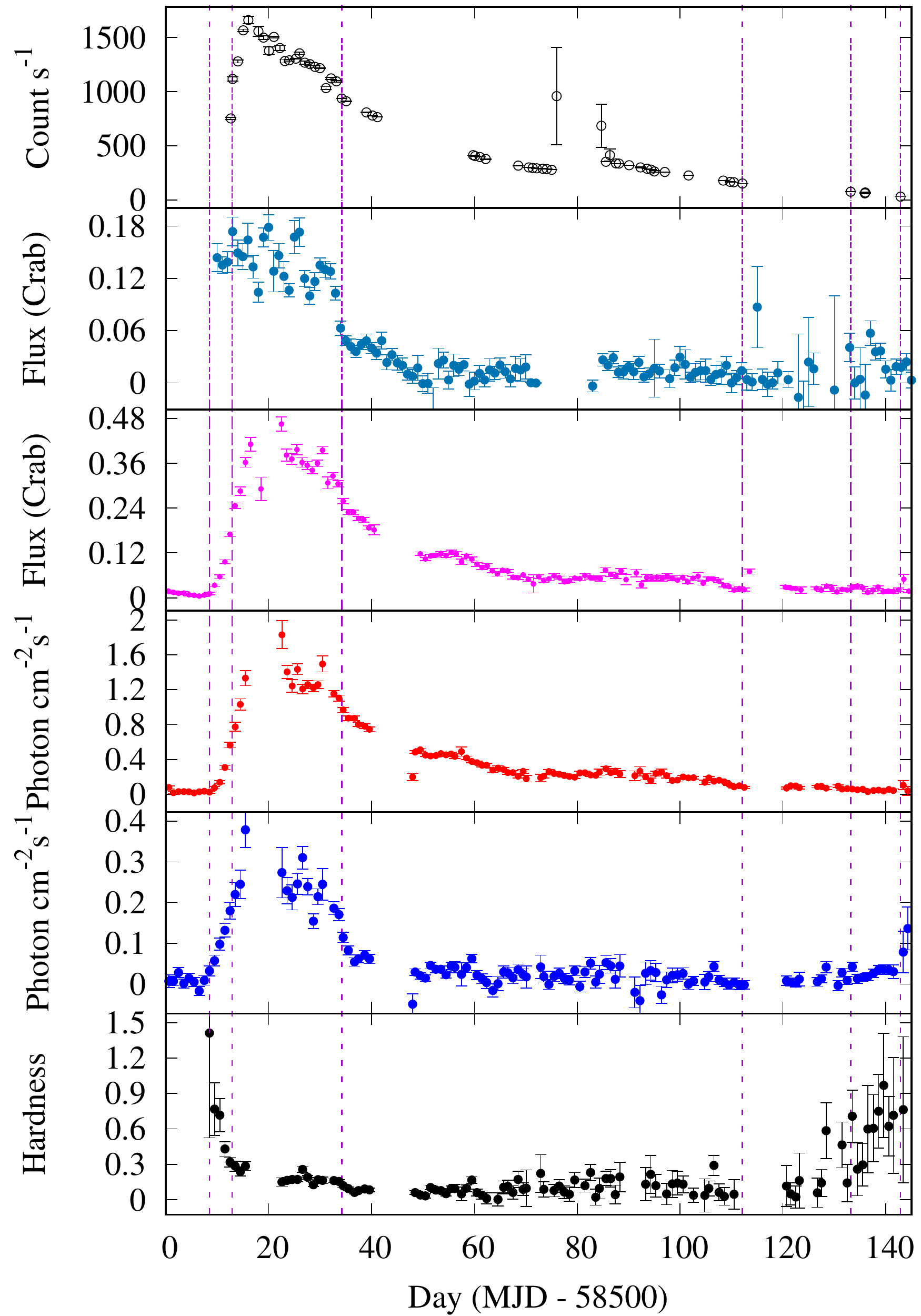}
\caption{Variation in flux and hardness ratio towards  Swift J1728.9--3613 observed by {\it NICER}, {\it Swift}/BAT and {\it MAXI}/GSC during the January 2019 outburst. The first, second, and third panels show the evolution of X-ray flux observed by {\it NICER} (0.2--12 keV), {\it Swift}/BAT (15--50 keV), and {\it MAXI}/GSC (2--20 keV), respectively between MJD 58500 and MJD 58645. We observed that the flux decreased very rapidly in the hard energy band (6--20 keV) compared to that in the soft energy band (2--6 keV) of {\it MAXI}/GSC, which are shown in the fourth and fifth panels respectively. The corresponding evolution of the hardness ratio (10--20 keV/2--4 keV) by {\it MAXI} is shown in the bottom (sixth) panel. The source experienced different state transitions during the outburst, which are indicated by vertical dotted lines.}
\label{fig:light}
}
\end{figure}

\subsection{Temporal variation of flux and hardness ratio}
\label{subsec:light_curve}
The temporal variation of flux towards Swift J1728.9--3613 during the January 2019 outburst as measured by {\it MAXI}/GSC\footnote{\href{http://maxi.riken.jp/top/index.html}{http://maxi.riken.jp/top/index.html}} and {\it Swift}/BAT\footnote{\href{https://swift.gsfc.nasa.gov/results/transients/}{https://swift.gsfc.nasa.gov/results/transients/}} was shown in Fig. \ref{fig:light}. In the first panel, the variation of flux obtained from {\it NICER} is shown. The second and third panels show fluxes observed by {\it Swift}/BAT (15--50 keV) and {\it MAXI}/GSC (2--20 keV) respectively. In soft energy bands the outburst started on MJD 58508.50 as seen from the MAXI (2-20 keV) light curve and reached the maximum flux within 15 days as we observe in the third panel of Fig. \ref{fig:light}. When Swift/BAT started its observation (on MJD 58511.05), the source reached close to its peak flux as observed from the BAT light curve. It is clear that the outburst was started before MJD 58508.50.
The outburst was characterized by a sharp rise to the peak flux of $1.77\pm0.07$ photons cm$^{-2}$ s$^{-1}$ ($0.47\pm0.02$ Crab) on MJD 58522.50, followed by a very slow decay over $\sim$120 days until it became undetectable by {\it MAXI}/GSC. The source was detected by {\it Swift}/BAT on MJD 58511.05, but at the time of its discovery, the source reached near the peak as shown in the second panel of Fig. \ref{fig:light}. The peak flux obtained by {\it Swift}/BAT was $0.039\pm0.003$ count cm$^{-2}$ s$^{-1}$ ($0.18\pm0.02$ Crab) on MJD 58520 in the energy range of 15--50 keV. The source flux decreases faster in the hard X-ray band (BAT) compared to the soft band (MAXI), as seen in Fig. \ref{fig:light}. The source flux (15–50 keV) returned to a consistent level after the MJD 58550, as visible from the BAT light curve. The soft X-ray energy band (MAXI) light curve indicates that the source returns to a consistent level about $\sim$20 days later (on MJD 58570) compared to the hard X-ray band. 

In the lower three panels of Fig. \ref{fig:light}, the variation of fluxes in different energy bands of {\it MAXI}/GSC as well as the variation of HR with time (MJD) are shown. The fourth and fifth panels show the variation of fluxes in the energy ranges of 2--6 keV and 6--20 keV of {\it MAXI}/GSC.  We observed that the flux rapidly decreased to a sufficiently lower value in the 6--20 keV energy band after MJD 58550 which we also observed in the BAT light curve (second panel). The soft band flux (in 2--6 keV of GSC) rather decayed at a much lower rate, and the outburst continued up to approximately MJD 58642.50. The sixth panel shows the evolution of HR (10--20 keV/2--4 keV) during the outburst. Initially, when the source was detected on MJD 58511.05, it was found in the HS (HR was as high as $\sim$1.4), and in the next observation on MJD 58512.64, it was found in the SIMS as observed by {\it NICER} (this state classification is segregate based upon the study of PDS and energy spectra). The source remained in SIMS for $\sim$22 days (up to MJD 58533.17), after which it was found in the SS on MJD 58534.20 (from the study of PDS and energy spectra). After remaining in the SS for almost 78 days, HR started to increase, and the source started the soft-to-hard transition. The study of PDS and energy spectra suggests that the source was in the HIMS on MJD 58633.17. On the next {\it NICER} observation on MJD 58642.88, the source was found in the HS at the end of the outburst. Due to less frequent observations at the decaying phase, the exact time of transitions was not found.

\begin{figure*}
\centering{
\includegraphics[width=6cm,angle=270]{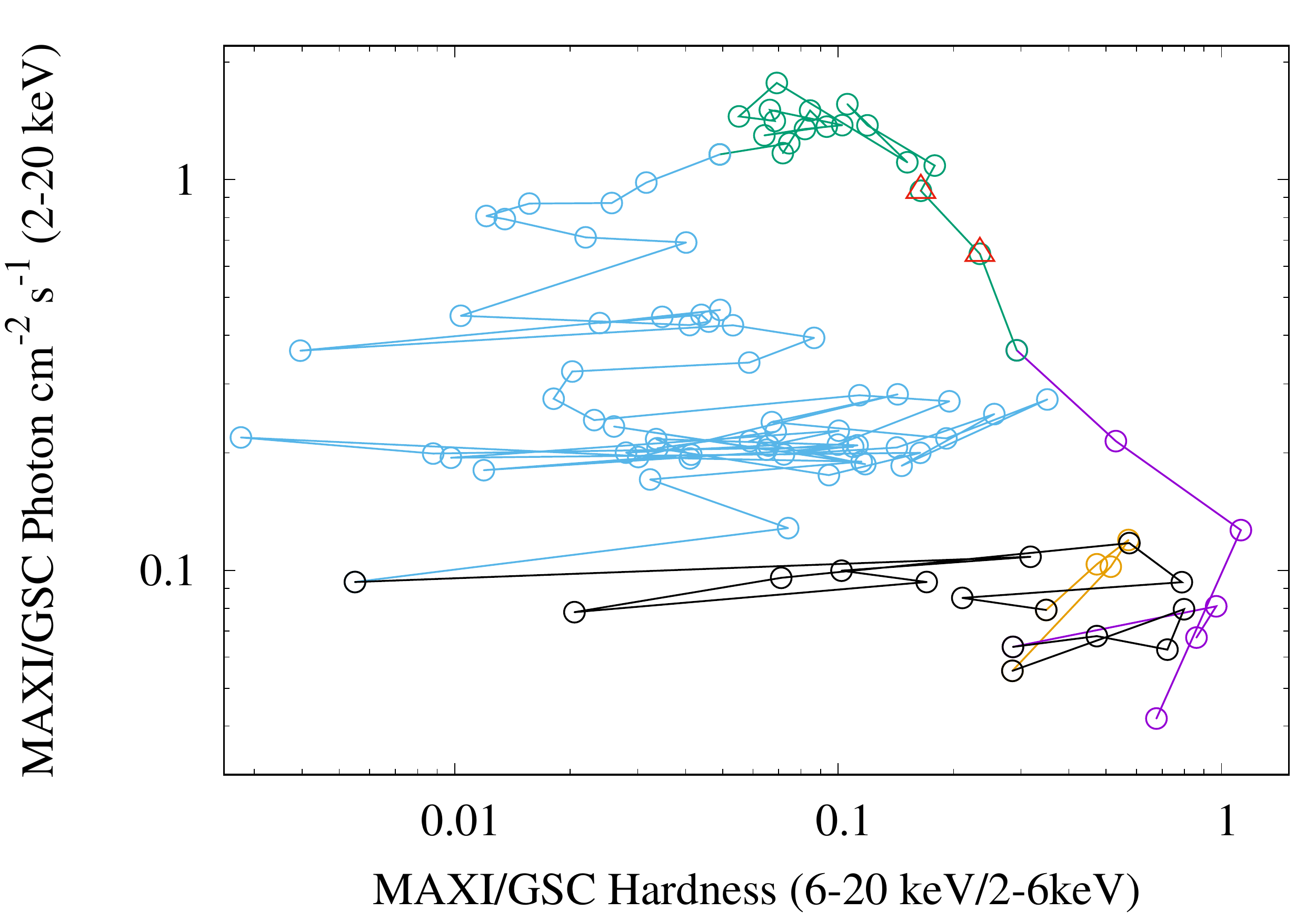}
\includegraphics[width=6cm,angle=270]{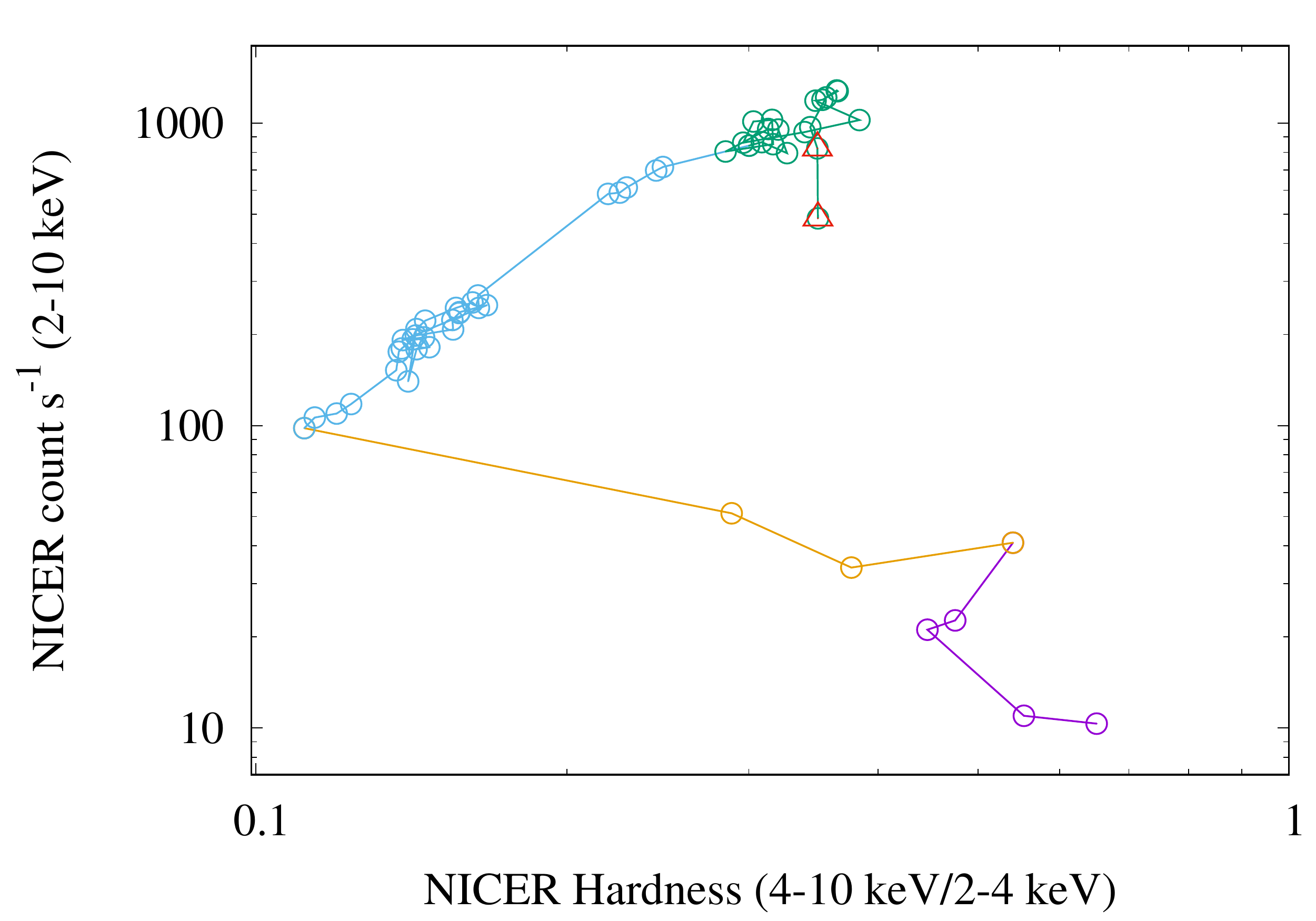}
\caption{HID of Swift J1728.9--3613 during January 2019 outburst using {\it MAXI}/GSC (left) and {\it NICER} (right). The horizontal axis of the left figure shows the variation of hardness ratio (10--20 keV/2--4 keV) and the vertical axis indicates the 2--20 keV photon flux of the source as observed by {\it MAXI}/GSC. In the right-hand side figure, flux in 2--10 keV of {\it NICER} is plotted with respect to 4--10 keV/2--4 keV {\it NICER} hardness ratio. Spectral states are denoted by using different colors (violet: HS; green: SIMS; cyan: SS; orange: HIMS). Black points are those which remained unidentified towards the end of the outburst due to the lack of {\it NICER} observation. Two red triangles indicate the positions of type B QPOs found  during the outburst.}
\label{fig:hid}
}
\end{figure*}

Evolution of the source through different states is shown in the HID (Fig. \ref{fig:hid}), where the {\it MAXI}/GSC hardness (10--20 keV/2--4 keV) is plotted with the 2--20 keV flux of {\it MAXI}/GSC (left). We have also shown the HID using NICER data, in which the hardness ratio (4--10 keV/2--4 keV) is plotted with respect to 2--10 keV count rate (right). The evolution started at MJD 58508.5 when the source was in the HS, but it underwent a rapid state transition with increasing intensity. The top region of the HID represents the SIMS, where the source stayed for almost 22 days before reaching the SS. The SS is indicated by the left branch in the HID. During the SS, the HR was highly fluctuating. On MJD 58633.17, the source reached the HIMS and finally reached the HS on MJD 58642.88. The complete evolution made a ``$q$''-shaped track in the HID traversing in the anticlockwise direction, which is typically observed in black hole X-ray binaries \citep{Be05, Ho05}.

\subsection{Evolution of fractional RMS}
\label{subsec:variability}

\begin{figure}
\includegraphics[width=6.2cm,angle=270]{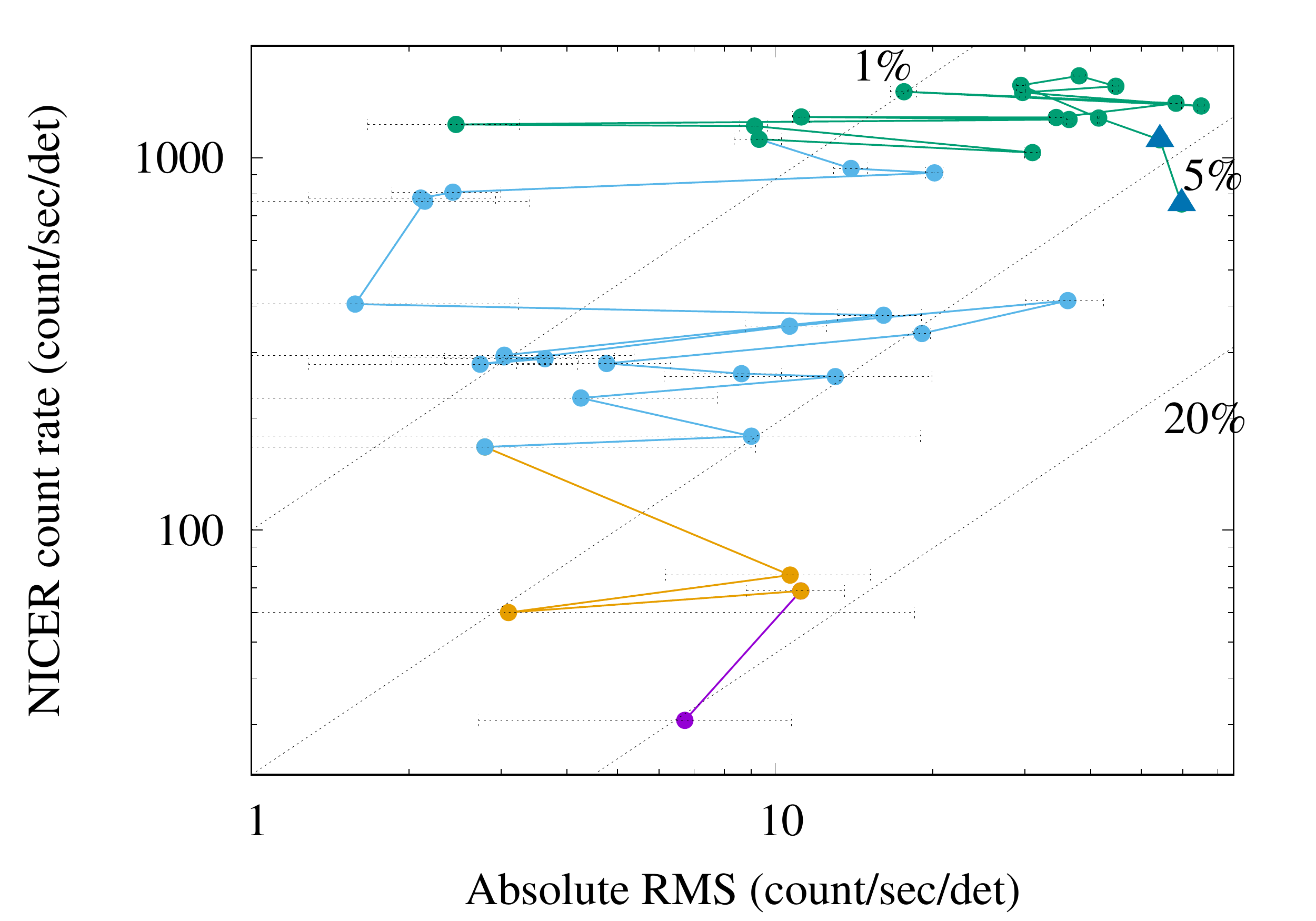}
\caption{Figure shows the RMS-intensity diagram during the January 2019 outburst of Swift J1728.9--3613 where the absolute RMS (using {\it NICER} data) is plotted against the total count rate in the 0.2--12 keV energy range of {\it NICER}. A partial hysteresis is present in the RID, which is typically observed when a black hole undergoes state transition during outbursts. Spectral states are denoted by using different colors (violet: HS; green: SIMS; cyan: SS; orange: HIMS). Blue solid triangles indicate the positions of type B QPOs found  during the outburst.}
\label{fig:hrd_rid}
\end{figure}

In Fig. \ref{fig:hrd_rid}, we have shown the evolution of the source in the RID where the total absolute RMS was plotted with the {\it NICER} count rate during the outburst. The total absolute RMS was calculated by multiplying the fractional RMS (using {\it NICER} data) with the {\it NICER} count rate \citep{Mu11}. The PDS were studied using {\it powspec} using the light curves in 0.2--12 keV energy range with bin time of 0.01 s. The {\it NICER} observation started when the PDS was dominated by the weak broadband noise component (fractional RMS was approximately 8 per cent) with a net count rate of $752.1\pm9.0$ count s$^{-1}$ det$^{-1}$ on MJD 58512.64, and the source had already left the hard-line (HL; see Fig. \ref{fig:hrd_rid}). This result corresponds to the SIMS of the black hole which was also reported by \citet{En19}. We could not study the evolution of the source before MJD 58512.64 since no {\it NICER} observation was performed before that day and the source evolved very rapidly during that period. On MJD 58534.20, the fractional RMS decreased to $\sim$1, and the source was found in SS (the spectral study also suggests the same). The intensity reached a peak value of $1661.7\pm32.5$ count s$^{-1}$ det$^{-1}$ on MJD 58516.05 followed by a gradual decrease over the next 125 days approximately. RMS changed from $\sim$1.67 per cent on MJD 58609.77 to $\sim$14.11 per cent on MJD 58633.17 which suggests a transition from soft to hard states during this period, but the exact time of transition was not found because of any {\it NICER} observation during that time. Fractional RMS reached $\sim$22 per cent on MJD 58642.88 and the {\it NICER} count rate also decreased to $30.8\pm0.1$ count s$^{-1}$ det$^{-1}$. The source seemed to return toward the adjacent hard line (AHL) towards the end of the outburst as seen from Fig. \ref{fig:hrd_rid}. The whole evolution on RID showed a partial hysteresis, which is a typical phenomenon to be observed in black hole X-ray binaries \citep{Mu11}.

\begin{figure*}
\centering{
\includegraphics[width=5.6cm,angle=270]{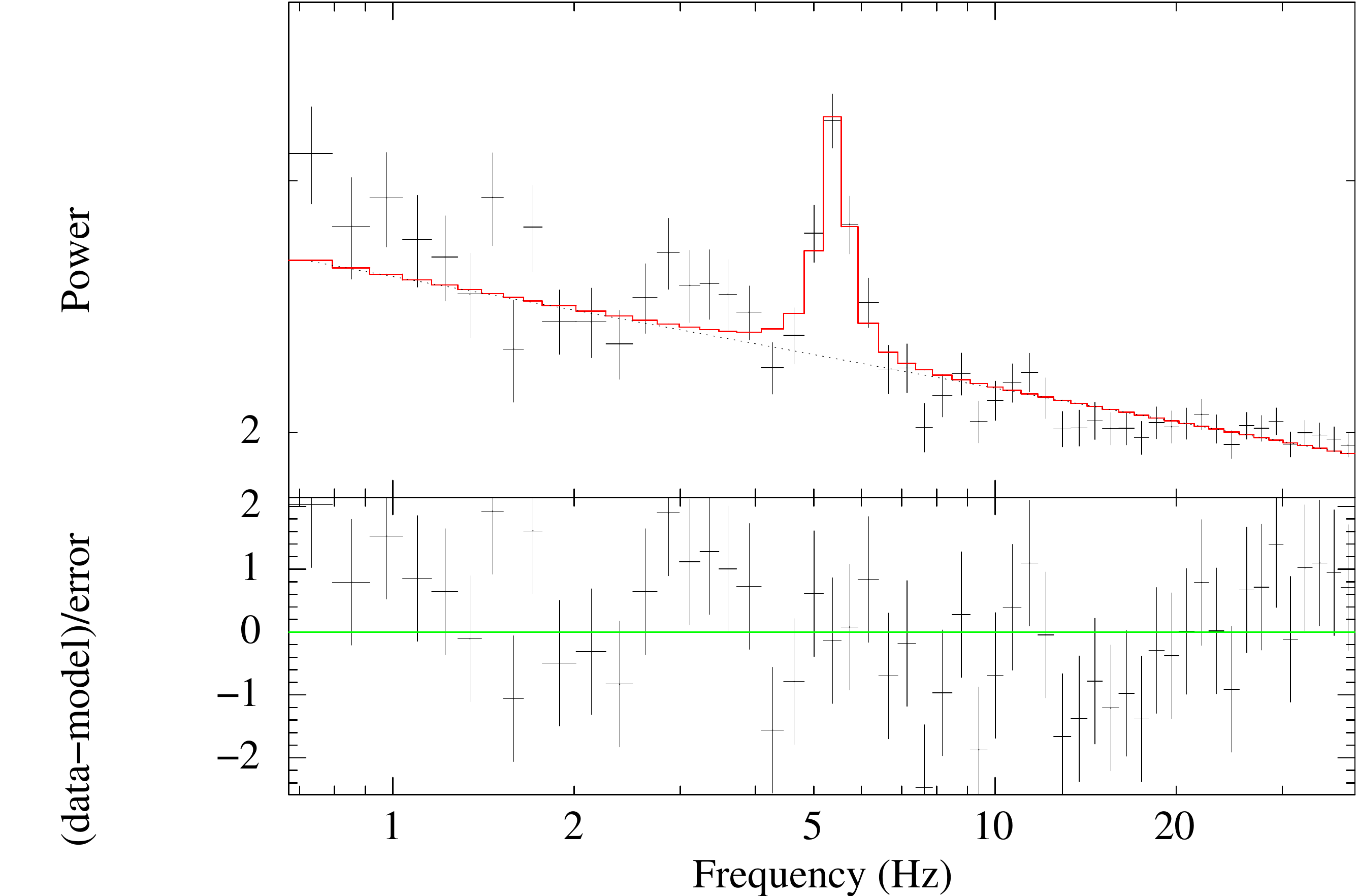}
\includegraphics[width=5.75cm,angle=270]{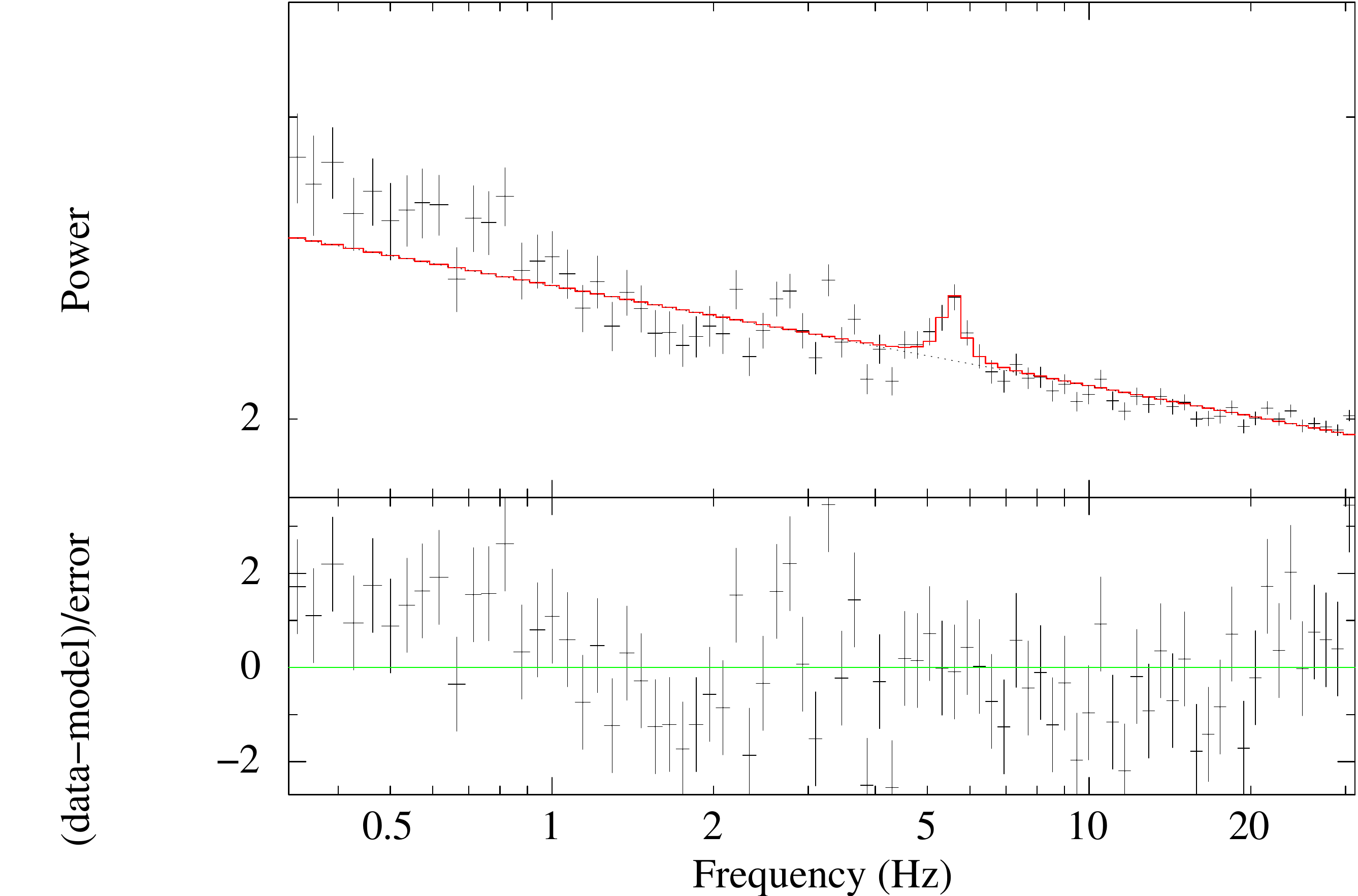}
	\caption{The power density spectra showing two QPOs during the outburst. The left-hand side and right-hand side figures show the QPOs found on MJD 58512.64 and MJD 58513.02 at 5.40$\pm$0.05 Hz and 5.55$\pm$0.06 Hz respectively when the source was in the SIMS. QPOs are identified as type-B QPOs. The PDS is fitted using a power-law and a Lorentzian component. The fitting parameters are summarized in Table \ref{tab:qpo}}.
\label{fig:pds}
	}
\end{figure*} 

\subsection{Study of power density spectra}
\label{subsec:pds}

We study the PDS and the corresponding evolution of power spectral properties of the new transient during the entire phase of the 2019 outburst. Leahy-normalized PDS \citep{Le83} are produced using 0.01 s binned light curves for NICER observations using the {\tt FTOOL} task {\tt POWSPEC} in {\tt HEASOFT}. Light curves were divided into stretches of 8192 bins per interval. The final PDSs for each observation are produced from the average of PDSs from all of the segments. During the outburst (from MJD 58512.64 to MJD 58657.00), two QPOs are found. The PDS is fitted with two components, a power law, and a Lorentzian to fit the QPO peaks. We have fitted the power spectrum in XSPEC (details can be found in \citep{In12}) and uncertainties are given for a 90 per cent confidence interval. PDS is produced using NICER data that have not been corrected for dead time. The dead-time has an impact on the PDS's white noise contribution and occasionally the QPO's overall form, but it has no impact on the peak frequency of the QPO. The Q-factor and significance are impacted by a distortion of the overall QPO shape, however, the variation is only obvious at higher count rates (above 20,000 NICER counts s$^{-1}$ using 52 FPM)\footnote{\href{https://heasarc.gsfc.nasa.gov/docs/nicer/data\_analysis/nicer\_analysis\_tips.html}{https://heasarc.gsfc.nasa.gov/docs/nicer/data\_analysis/nicer\_analysis\_tips.html}}, above 600 counts s$^{-1}$ for NuSTAR \citep{Ba15}. In Fig. \ref{fig:pds}, we have shown different QPOs detected during the January 2019 outburst of Swift J1728.9--3613. The corresponding fitting parameters of various QPOs are shown in Table \ref{tab:qpo}. Quality factors were calculated using the ratio of QPO frequency to the FWHM (i.e., $Q=\nu/\Delta\nu$). The significance of the QPOs is estimated using the ratio of normalization to the negative error of the normalization of the Lorentzian component \citep{Al14, Ba15}.

On the first three {\it NICER} observations, PDS was characterized by a broadband noise component with fractional RMS varying between $\sim$8 per cent and $\sim$4 per cent, and two significant narrow and strong QPOs were detected (see Tab. \ref{tab:qpo}). The top panel of Fig. \ref{fig:pds} showed two significant QPOs found on MJD 58512.64 (6.68$\sigma$) and MJD 58513.02 (3.86$\sigma$), respectively. The QPO frequencies increased from $5.40\pm0.05$ Hz to $5.56\pm0.06$ Hz and the corresponding RMS amplitude decreased from $4.30\pm0.46$ per cent to $1.08\pm0.21$ per cent. These parameters (such as centroid frequencies, RMS amplitude, significance) are well consistent with those found in type-B QPOs (for detail, Table 1 of \citep{Ca05}), which are typically found in the SIMS. It should be noted that the QPO obtained on MJD 58512.64 was also found previously by \citet{En19} during its initial observations by {\it NICER}.

\begin{table}
\centering
\caption{Summary of the QPOs detected during different phases of the January 2019 outburst of Swift J1728.9--3613.}
\label{tab:qpo}
\begin{tabular}{cccccc}
\hline
MJD	           	& Frequency	            & $Q$-value     	    & RMS               & $\sigma$	& QPO	    \\
	        		        & (Hz)		            &			            & (per cent)	    &           & type      \\
\hline
58512.64			& $5.40\pm0.05$		    & $7.78\pm0.19$			& $4.30\pm0.46$		& 6.68      & B	        \\
58513.02			& $5.56\pm0.06$		    & $11.49\pm0.37$		& $1.08\pm0.21$		& 3.86      & B	        \\
\hline

\end{tabular}
\end{table}

\subsection{Evolution of spectral parameters}
\label{subsec:spectrum}
The study of timing properties showed that the new source Swift J1728.9--3613 shows properties that are similar to black hole X-ray binaries. Based on this realization, we further studied the energy spectra of the source during different phases of the outburst using {\it Swift}/XRT and {\it NICER} data.
{\it Swift} performed the first observation on the day of its discovery (MJD 58511.05, Obs Id -- 00886157001). {\it Swift}/XRT energy spectra (1--10 keV) are fitted by using the model {\it tbabs$\times$(nthcomp+diskbb)} (same as used to fit the {\it NICER} spectra also) which results in reduced $\chi^{2}=1.04$ for 899 degrees of freedom. During this period, the spectral index ($\Gamma$) was $1.80\pm0.32$ with an inner disc temperature of $0.785\pm0.541$ keV, indicating that the source was in the HS on MJD 58511.05 during the rising phase of the outburst. The disc normalization is significantly high compared to that observed in the SS. Flux obtain from the XRT spectrum remains consistent with the fluxes obtained from the {\it NICER} spectra.

{\it NICER} observations were started at MJD 58512.64 (the day after discovery). {\it NICER} monitored the source regularly and performed a total $\sim$175 ks of observation between MJD 58512.64 and MJD 58657.
Energy spectra are studied in the energy range of 1--10 keV using the X-ray spectral fitting package {\tt XSPEC v12.10.1} \citep{Ar96}. Each spectrum is grouped to get at least 25 counts for each bin using the task {\tt grppha}. Energy spectra are fitted with a two-component model consisting of a thermally comptonized continuum ({\it nthccomp}: \citet{Zd96, Zy99}) and a multi-temperature disk blackbody ({\it diskbb}: \citet{Mi84, Ma86}) components. We have used {\it tbabs} \citep{Wi00} to model the absorption along the source direction. We look for the presence of an iron emission line near $\sim$6.4 keV by adding a {\it Gaussian} component but did not find its evidence in any of the {\ NICER} observations. The unabsorbed flux of different components of the model is estimated by using {\it cflux}\footnote{\href{https://heasarc.gsfc.nasa.gov/xanadu/xspec/manual/node284.html}{https://heasarc.gsfc.nasa.gov/xanadu/xspec/manual/node284.html}} in the energy range of 1--10 keV.
During the fitting, we keep all parameters free to vary for all observations. The values of column densities are found to vary between 2.8$\times$10$^{22}$ cm$^{-2}$ and 3.7$\times$10$^{22}$ cm$^{-2}$ which is comparable to the previously obtained value by \citet{En19}. Spectral parameters obtained by fitting XRT and {\it NICER} spectra are given in Table \ref{tab:parameters}.

\begin{figure}
\centering{
\includegraphics[width=8.5cm]{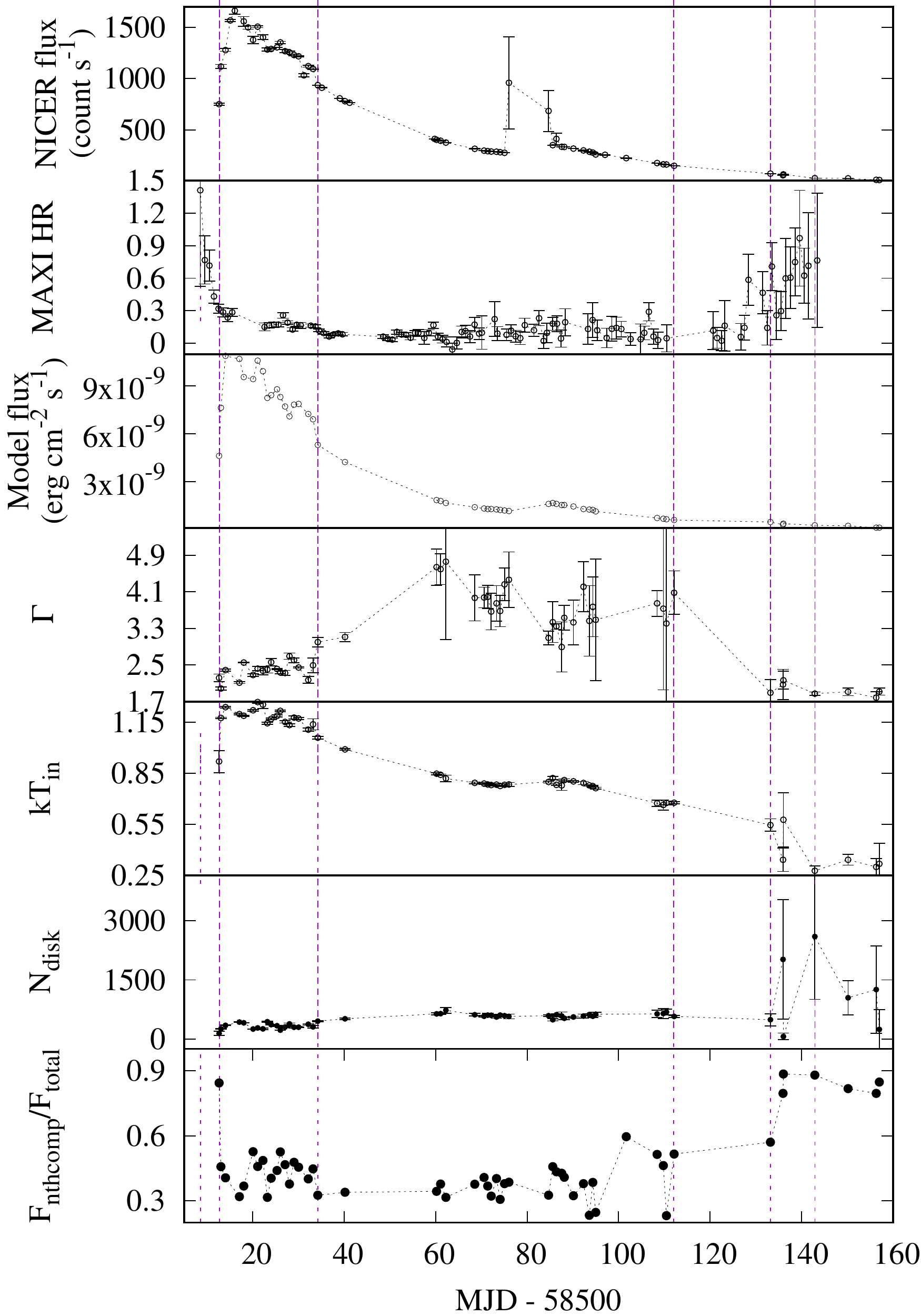}
\caption{The evolution of spectral parameters with time (MJD) during the outbursting phase. In the upper two panels, the variations of {\it NICER} flux (0.2--12 keV) and the variation of hardness ratio (10--20 keV/2--4 keV of {\it MAXI}) with time are shown. In the third panel, we have shown the variation of 1--10 keV flux was obtained from fitting the {\it NICER} spectra. The evolution of spectral index, inner disc temperature, and diskbb normalization are shown in the fourth, fifth, and sixth panels, respectively. In the bottom panel, we have shown the ratio of thermally comptonized hard photons flux to the total flux. Vertical lines indicate the time of transition through different spectral states during the outburst.}
\label{fig:parameters}
}
\end{figure}

The evolution of different spectral parameters during the outburst is shown in Fig. \ref{fig:parameters}. The variation of the {\it NICER} count rate and the {\it MAXI}/GSC hardness ratio (10--20 keV/2--4 keV) are shown in the first and second panels, respectively. The variation of average flux in the energy range 1--10 keV obtained from the spectral fitting is plotted in the third panel. In both panels 1 and 3, it is observed that the flux rapidly increased to a maximum on MJD 58521.05 followed by an exponential decay over several months. In the fourth panel, we have shown the evolution of the spectral index ($\Gamma$) during the outburst. The spectral index varied between $2.22\pm0.08$ and $2.69\pm0.07$ between MJD 58512.64 and MJD 58533.17, which indicated that the source remained in the intermediate states. The spectral index suddenly increases to $3.00\pm0.11$ on MJD 58534.20, which suggests a transition to the SS. A sharp change is observed in the timing evolution of the spectral index as well as in the evolution of {\it NICER} flux, hardness ratio, and model flux also. The spectral index is varying between $2.93\pm0.03$ and $4.76\pm1.71$ from MJD 58534.20 to MJD 58612.13 and decreases to $1.89\pm0.29$ on MJD 58633.17 which indicates a transition from SS to the intermediate states.
In the fifth and sixth panels of Fig. \ref{fig:parameters}, the evolution of inner disk temperature ($kT_{\text{in}}$) and the disk normalisation ($N{\text{disk}}$) are shown. The temperature at the beginning of the observation was $0.918\pm0.065$ keV on MJD 58512.64. Temperature raised to the maximum value of $1.267\pm0.006$ on MJD at 58521.05 when the flux was also maximum. The sixth panel shows that the disk normalization remained almost constant between MJD 58534.20 and 58612.13 (indication of ISCO). 
The disc temperature remained almost constant between MJD 58568.53 and MJD 58601.65, which slightly varied within the range from $0.762\pm0.006$ keV to $0.821\pm0.009$ keV.

In the bottom panel (seventh) of Fig. \ref{fig:parameters}, shows the ratio of flux due to the thermally comptonized photons to the total flux ($F_\text{nthcomp}/F_\text{total}$) during the outburst. We observe that at the beginning, flux due to comptonized photons was about 80\% of the total flux. The ratio seemed to vary between 30 per cent  and 60 per cent  between MJD 58513.02 and MJD 58533.17. During MJD 58534.20 to MJD 58612.13, when the source was in SS, the flux due to thermal comptonization was as minimum as 30--40\%, except for some increase towards the end of the SS. The evolution of spectral parameters indicates that spectra are dominated by the disk photons during the SS, but at the end of the SS, an effect due to disk photons became less. After SS, the spectra were again dominated by thermally comptonized hard photons that contributed to more than 80 per cent  of the total flux.

We observed that the {\it NICER} flux (in the first panel) suddenly increased from $\sim$279 count s$^{-1}$ on MJD 58575.05 to $\sim$959 count s$^{-1}$ on MJD 58576.01 within one-day time-scale. The flux remained high till the next {\it NICER} observation also, which was performed on MJD 58584.69. But, this probably happened due to solar activity. Spectral parameters are also found to vary during this period. Model flux obtained from the spectral fitting was found to increase and the corresponding spectral index significantly decreased during this period. The spectral index suddenly decreased from $4.36\pm0.61$ on MJD 58576.01 to $3.09\pm0.15$ on MJD 58584.69 indicating that the spectra became slightly harder during that event. The inner disc temperature increased from $0.783\pm0.012$ on MJD 58576.01 to $0.798\pm0.006$ on MJD 58584.69, which further raised to a higher value of $0.821\pm0.009$ on MJD 58585.61, as shown in the fifth panel of Fig. \ref{fig:parameters}. Disk normalization ($N_\text{disk}$), which is a measure of the radius of ISCO, does not vary much during this period.

\begin{table*}
\centering
\caption{Fitting parameters of the source for the model {\it tbabs$\times$(nthcomp+diskbb)} using {\it NICER} and XRT data.}
\label{tab:parameters}
\begin{tabular}{lccccccccr}
\hline
Instrument & MJD & Obs Id & Exposure & $n_{\text{H}}$    & $\Gamma$ & $kT_{\text{in}}$     & $N_{\text{disk}}$   & Model flux      & $\chi^{2}$/d.o.f.   \\
           &     &           & (ks)   & (10$^{22}$ cm$^{-2}$)  &    & (keV)                &                      & (erg cm$^{-2}$ s$^{-1}$)   &          \\
\hline
XRT & 58511.05 & 00886157001 & 0.995 & 3.05$\pm$0.31   & 1.80$\pm$0.32 & 0.785$\pm$0.541 & 2157.3$\pm$4106.5 & 5.49$\times$10$^{-10}$ & 938/899  \\
\hline
& 58512.64 & 1200550101 & 1.273 & 3.50$\pm$0.05 & 2.22$\pm$0.08 & 0.918$\pm$0.065 & 144.9$\pm$57.9& 4.61$\times$10$^{-9}$ & 975/815  \\
& 58513.02 & 1200550102 & 11.716& 3.35$\pm$0.01 & 1.98$\pm$0.04 & 1.175$\pm$0.004 & 261.3$\pm$4.5 & 7.64$\times$10$^{-9}$ & 3074/899 \\
& 58514.03 & 1200550103 & 14.070& 3.57$\pm$0.01 & 2.39$\pm$0.02 & 1.237$\pm$0.005 & 348.0$\pm$8.7 & 1.09$\times$10$^{-8}$ & 4282/899 \\
& 58517.08 & 1200550106 & 19.124& 3.41$\pm$0.01 & 2.11$\pm$0.02 & 1.197$\pm$0.004 & 431.3$\pm$4.5 & 1.07$\times$10$^{-8}$ & 6835/898 \\
& 58518.02 & 1200550107 & 10.744& 3.56$\pm$0.01 & 2.55$\pm$0.01 & 1.187$\pm$0.003 & 411.7$\pm$5.3 & 9.57$\times$10$^{-9}$ & 5246/898 \\
& 58520.08 & 1200550109 & 7.382 & 3.45$\pm$0.01 & 2.28$\pm$0.02 & 1.219$\pm$0.003 & 253.8$\pm$7.5 & 9.44$\times$10$^{-9}$ & 2870/899 \\
& 58521.05 & 1200550110 & 3.360 & 3.56$\pm$0.02 & 2.42$\pm$0.05 & 1.267$\pm$0.006 & 281.0$\pm$7.3 & 1.06$\times$10$^{-8}$ & 2290/899 \\
& 58522.22 & 1200550111 & 1.482 & 3.53$\pm$0.03 & 2.37$\pm$0.08 & 1.253$\pm$0.022 & 259.5$\pm$20.0 & 9.93$\times$10$^{-9}$ & 1439/869 \\
& 58523.18 & 1200550112 & 1.369 & 3.42$\pm$0.03 & 2.40$\pm$0.07 & 1.143$\pm$0.006 & 438.5$\pm$15.6 & 8.26$\times$10$^{-9}$ & 1335/836 \\
& 58524.03 & 1200550113 & 5.000 & 3.53$\pm$0.03 & 2.56$\pm$0.08 & 1.169$\pm$0.004 & 366.8$\pm$7.2 & 8.43$\times$10$^{-9}$ & 2879/898 \\
& 58525.31 & 1200550114 & 4.209 & 3.48$\pm$0.01 & 2.41$\pm$0.01 & 1.186$\pm$0.006 & 336.2$\pm$12.9 & 8.80$\times$10$^{-9}$ & 2434/898 \\
& 58526.02 & 1200550115 & 8.023 & 3.47$\pm$0.01 & 2.34$\pm$0.03 & 1.216$\pm$0.004 & 227.1$\pm$4.8 & 8.31$\times$10$^{-9}$ & 3746/899 \\
& 58527.05 & 1200550116 & 4.471 & 3.43$\pm$0.02 & 2.32$\pm$0.05 & 1.151$\pm$0.007 & 304.3$\pm$11.5 & 7.72$\times$10$^{-9}$ & 2346/897 \\
& 58528.02 & 1200550117 & 1.061 & 3.52$\pm$0.05 & 2.69$\pm$0.07 & 1.132$\pm$0.008 & 386.0$\pm$20.8 & 7.10$\times$10$^{-9}$ & 1318/800 \\
& 58529.03 & 1200550118 & 1.896 & 3.56$\pm$0.03 & 2.60$\pm$0.05 & 1.176$\pm$0.010 & 299.3$\pm$16.7 & 7.83$\times$10$^{-9}$ & 1742/872 \\
& 58530.00 & 1200550119 & 3.178 & 3.47$\pm$0.02 & 2.44$\pm$0.01 & 1.171$\pm$0.005 & 302.7$\pm$5.5 & 7.88$\times$10$^{-9}$ & 1661/893 \\
& 58532.14 & 1200550121 & 0.920 & 3.37$\pm$0.04 & 2.17$\pm$0.07 & 1.106$\pm$0.008 & 376.1$\pm$18.4 & 7.25$\times$10$^{-9}$ & 1050/817 \\
& 58533.17 & 1200550122 & 0.706 & 3.47$\pm$0.05 & 2.49$\pm$0.16 & 1.137$\pm$0.033 & 312.7$\pm$33.9 & 6.91$\times$10$^{-9}$ & 1016/778   \\
& 58534.20 & 1200550123 & 1.174 & 3.50$\pm$0.05 & 3.00$\pm$0.11 & 1.058$\pm$0.008 & 454.4$\pm$14.7 & 5.30$\times$10$^{-9}$ & 1240/762   \\
& 58535.10 & 1200550124 & 1.373 & 3.50$\pm$0.07 & 2.93$\pm$0.03 & 1.038$\pm$0.005 & 482.3$\pm$15.0 & 5.14$\times$10$^{-9}$ & 1407/772   \\
& 58539.02 & 1200550125 & 3.495 & 3.56$\pm$0.04 & 3.11$\pm$0.10 & 1.004$\pm$0.003 & 500.0$\pm$9.6 & 4.41$\times$10$^{-9}$ & 2161/823  \\
& 58540.18 & 1200550126 & 2.327 & 3.56$\pm$0.04 & 3.11$\pm$0.10 & 0.990$\pm$0.004 & 513.4$\pm$19.7 & 4.22$\times$10$^{-9}$ & 1451/784   \\
& 58541.15 & 1200550127 & 0.714 & 3.57$\pm$0.07 & 3.20$\pm$0.20 & 0.989$\pm$0.008 & 510.8$\pm$21.8 & 4.10$\times$10$^{-9}$ & 958/677   \\
& 58559.78 & 2200550103 & 0.468 & 3.59$\pm$0.27 & 3.89$\pm$0.99 & 0.847$\pm$0.008 & 607.1$\pm$24.0 & 1.90$\times$10$^{-9}$ & 558/538   \\
{\it NICER} & 58560.17 & 2200550104 & 1.227 & 3.71$\pm$0.12 & 4.64$\pm$0.40 & 0.847$\pm$0.006 & 639.0$\pm$24.7 & 1.83$\times$10$^{-9}$ & 701/613   \\
& 58561.07 & 2200550105 & 1.872 & 3.69$\pm$0.09 & 4.60$\pm$0.34 & 0.841$\pm$0.004 & 643.9$\pm$18.1 & 1.78$\times$10$^{-9}$ & 918/669   \\
& 58562.23 & 2200550106 & 0.356 & 3.72$\pm$0.28 & 4.76$\pm$1.71 & 0.819$\pm$0.019 & 729.7$\pm$73.3 & 1.65$\times$10$^{-9}$ & 527/500   \\
& 58568.53 & 2200550107 & 0.565 & 3.62$\pm$0.15 & 3.97$\pm$0.50 & 0.793$\pm$0.006 & 617.8$\pm$36.0 & 1.39$\times$10$^{-9}$ & 618/517   \\
& 58570.59 & 2200550108 & 1.697 & 3.65$\pm$0.07 & 3.97$\pm$0.22 & 0.789$\pm$0.004 & 578.8$\pm$17.4 & 1.31$\times$10$^{-9}$ & 875/607 \\
& 58571.36 & 2200550109 & 0.896 & 3.61$\pm$0.11 & 3.99$\pm$0.26 & 0.783$\pm$0.007 & 608.2$\pm$48.3 & 1.27$\times$10$^{-9}$ & 658/544   \\
& 58572.13 & 2200550110 & 2.445 & 3.48$\pm$0.10 & 3.67$\pm$0.40 & 0.781$\pm$0.003 & 601.3$\pm$22.5 & 1.27$\times$10$^{-9}$ & 958/635   \\
& 58573.30 & 2200550111 & 1.779 & 3.61$\pm$0.10 & 3.85$\pm$0.39 & 0.783$\pm$0.005 & 558.5$\pm$28.3 & 1.25$\times$10$^{-9}$ & 948/635   \\
& 58574.08 & 2200550112 & 3.050 & 3.47$\pm$0.09 & 3.68$\pm$0.34 & 0.775$\pm$0.003 & 608.6$\pm$15.3 & 1.22$\times$10$^{-9}$ & 1067/688 \\
& 58575.05 & 2200550113 & 1.803 & 3.63$\pm$0.11 & 4.26$\pm$0.36 & 0.781$\pm$0.005 & 591.3$\pm$28.4 & 1.18$\times$10$^{-9}$ & 824/607 \\
& 58576.01 & 2200550114 & 0.844 & 3.64$\pm$0.18 & 4.36$\pm$0.61 & 0.783$\pm$0.012 & 574.6$\pm$59.5 & 1.15$\times$10$^{-9}$ & 505/476   \\
& 58584.69 & 2200550115 & 2.273 & 3.39$\pm$0.06 & 3.09$\pm$0.15 & 0.798$\pm$0.006 & 591.1$\pm$22.2 & 1.60$\times$10$^{-9}$ & 1005/641   \\
& 58585.61 & 2200550116 & 0.770 & 3.59$\pm$0.14 & 3.44$\pm$0.45 & 0.821$\pm$0.009 & 489.6$\pm$32.1 & 1.65$\times$10$^{-9}$ & 672/551   \\
& 58586.38 & 2200550117 & 2.387 & 3.60$\pm$0.25    & 3.34$\pm$0.06 & 0.782$\pm$0.004 & 617.5$\pm$19.5 & 1.59$\times$10$^{-9}$ & 944/648   \\
& 58587.55 & 2200550118 & 0.457 & 3.52$\pm$0.16 & 2.89$\pm$0.54 & 0.779$\pm$0.030 & 584.8$\pm$100.3 & 1.52$\times$10$^{-9}$ & 645/491   \\
& 58588.12 & 2200550119 & 0.600 & 3.54$\pm$0.14 & 3.53$\pm$0.27 & 0.809$\pm$0.008 & 527.8$\pm$35.0 & 1.52$\times$10$^{-9}$ & 665/511  \\
& 58590.12 & 2200550120 & 1.436 & 3.43$\pm$0.10 & 3.43$\pm$0.49 & 0.801$\pm$0.004 & 558.8$\pm$22.5 & 1.43$\times$10$^{-9}$ & 848/588   \\
& 58592.30 & 2200550121 & 0.725 & 3.64$\pm$0.14 & 4.21$\pm$0.55 & 0.792$\pm$0.010 & 584.6$\pm$50.3 & 1.27$\times$10$^{-9}$ & 607/508   \\
& 58593.58 & 2200550122 & 1.379 & 3.35$\pm$0.20 & 3.46$\pm$0.77 & 0.778$\pm$0.004 & 619.8$\pm$21.9 & 1.24$\times$10$^{-9}$ & 697/562   \\
& 58594.36 & 2200550123 & 1.096 & 3.55$\pm$0.21 & 3.77$\pm$0.65 & 0.774$\pm$0.005 & 577.7$\pm$25.9 & 1.21$\times$10$^{-9}$ & 692/538   \\
& 58595.00 & 2200550124 & 1.555 & 3.38$\pm$0.28 & 3.49$\pm$1.33 & 0.762$\pm$0.006 & 628.0$\pm$50.7 & 1.12$\times$10$^{-9}$ & 723/565   \\
& 58608.40 & 2200550127 & 0.148 & 3.68$\pm$0.22 & 3.85$\pm$0.28 & 0.673$\pm$0.018 & 640.2$\pm$92.1 & 7.08$\times$10$^{-10}$& 339/302   \\
& 58609.77 & 2200550128 & 0.216 & 3.61$\pm$0.27 & 3.73$\pm$1.78 & 0.664$\pm$0.030 & 648.3$\pm$122.9 & 6.60$\times$10$^{-10}$& 337/329   \\
& 58610.45 & 2200550129 & 0.554 & 3.29$\pm$0.34 & 3.41$\pm$2.09 & 0.675$\pm$0.011 & 685.0$\pm$68.3 & 6.35$\times$10$^{-10}$& 465/407   \\
& 58612.13 & 2200550130 & 2.047 & 3.61$\pm$0.18 & 4.08$\pm$0.48 & 0.676$\pm$0.005 & 572.3$\pm$29.6 & 5.88$\times$10$^{-10}$& 657/520   \\
& 58633.17 & 2200550131 & 0.283 & 3.09$\pm$0.10 & 1.89$\pm$0.29 & 0.546$\pm$0.036 & 490.7$\pm$155.5 & 4.60$\times$10$^{-10}$ & 323/324   \\
& 58635.95 & 2200550132 & 0.151 & 3.21$\pm$0.16 & 2.07$\pm$0.33 & 0.342$\pm$0.069 & 2019.4$\pm$1515.3 & 2.94$\times$10$^{-10}$ & 190/174   \\
& 58636.01 & 2200550133 & 1.467 & 2.87$\pm$0.09 & 2.17$\pm$0.19 & 0.576$\pm$0.159 & 66.5$\pm$86.9 & 3.58$\times$10$^{-10}$ & 661/570   \\
& 58642.88 & 2200550134 & 0.987 & 3.09$\pm$0.20 & 1.87$\pm$0.04 & 0.276$\pm$0.029 & 2599.1$\pm$1595.1 & 2.46$\times$10$^{-10}$ & 433/428   \\
& 58650.17 & 2200550135 & 0.826 & 3.16$\pm$0.17 & 1.91$\pm$0.08 & 0.341$\pm$0.031 & 1046.4$\pm$436.8 & 2.12$\times$10$^{-10}$ & 350/383   \\
& 58656.31 & 2200550136 & 0.793 & 3.60$\pm$0.22 & 1.78$\pm$0.13 & 0.300$\pm$0.047 & 1257.0$\pm$1106.5 & 9.61$\times$10$^{-11}$ & 292/278   \\
& 58657.00 & 2200550137 & 1.757 & 3.71$\pm$0.26 & 1.91$\pm$0.08 & 0.317$\pm$0.122 & 244.1$\pm$506.2 & 1.07$\times$10$^{-10}$ & 582/385   \\
\hline
\end{tabular}
\end{table*}

\begin{figure*}
\centering{
\includegraphics[angle=270,width=8.5cm]{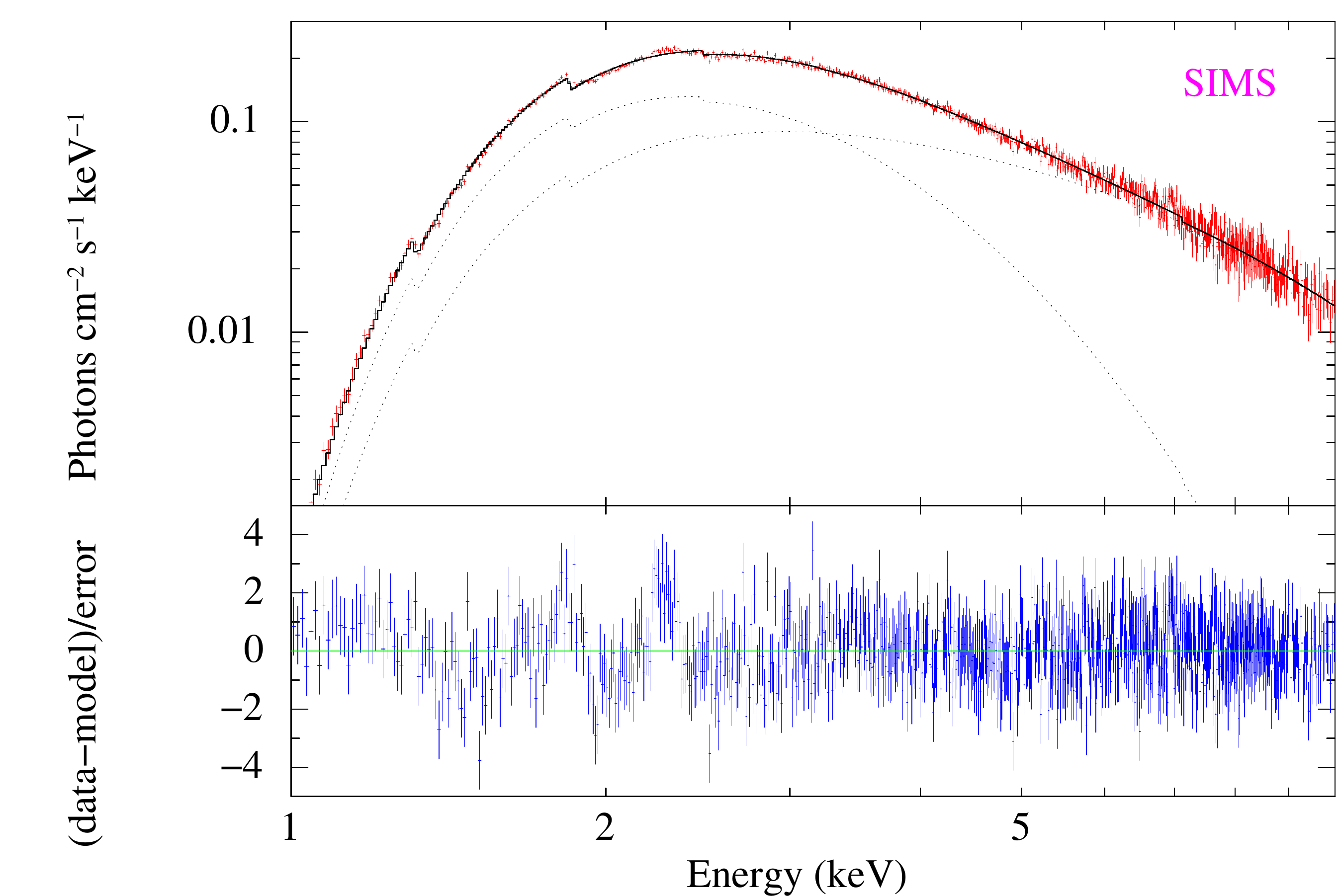}
\includegraphics[angle=270,width=8.5cm]{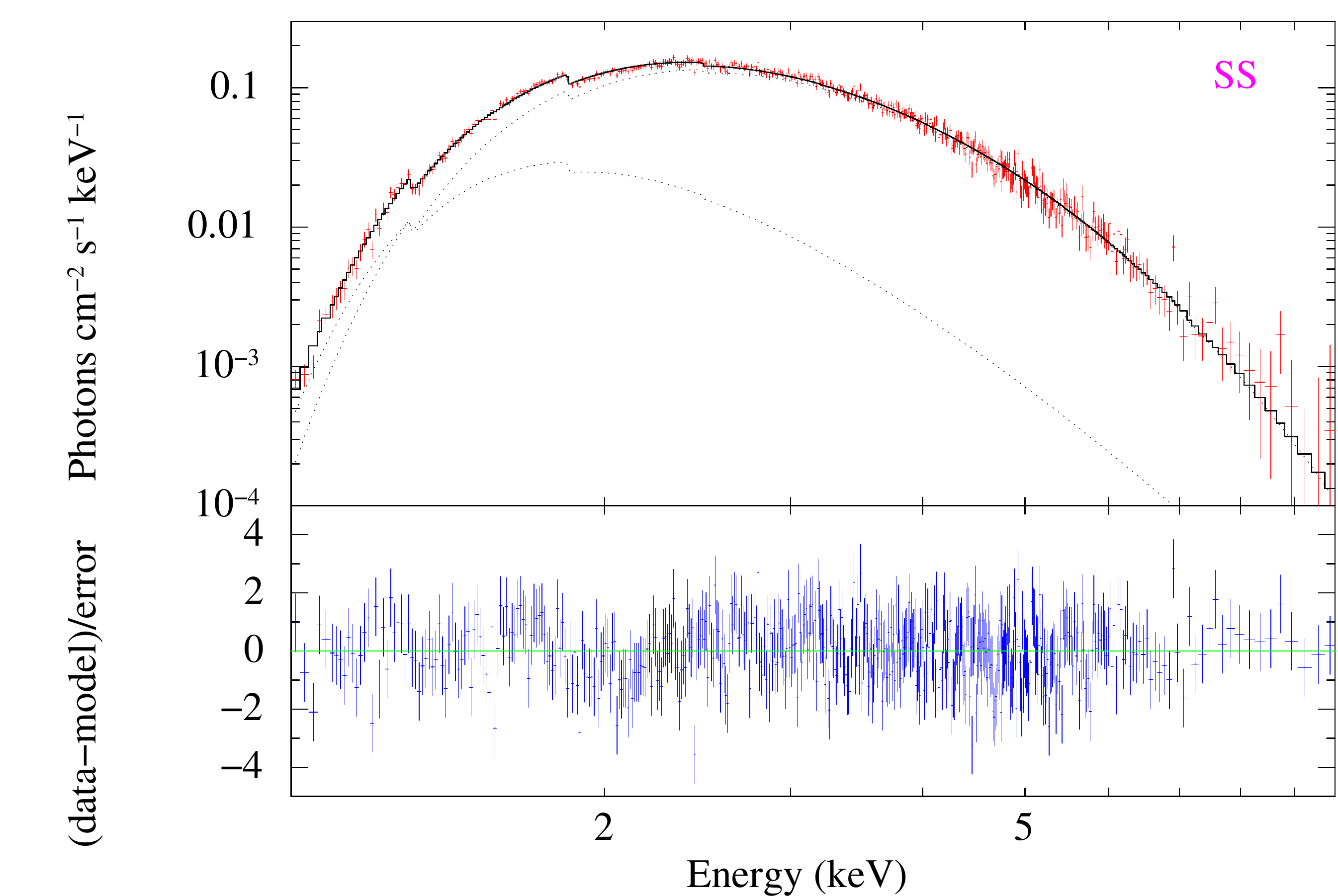}
\includegraphics[angle=270,width=8.5cm]{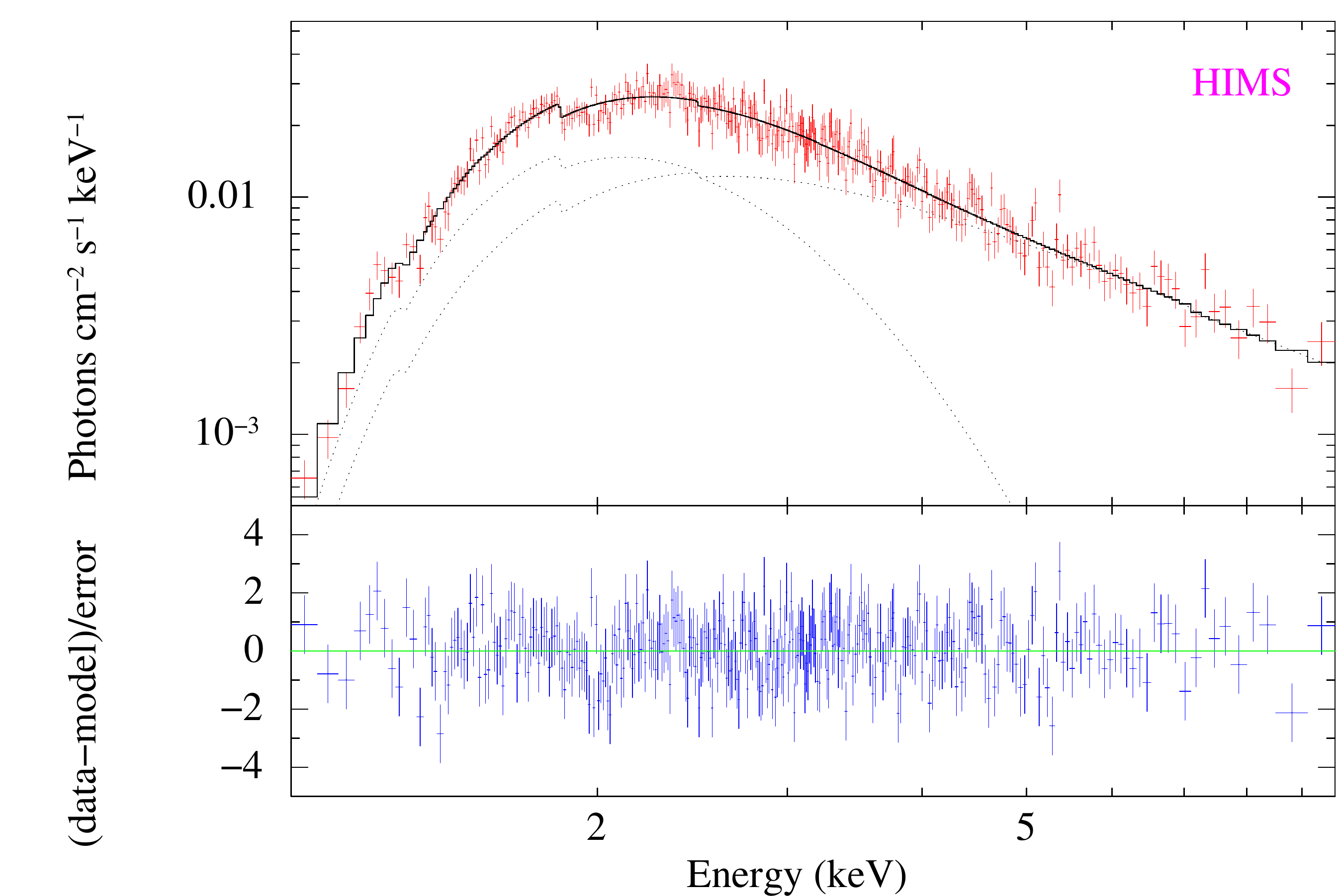}
\includegraphics[angle=270,width=8.5cm]{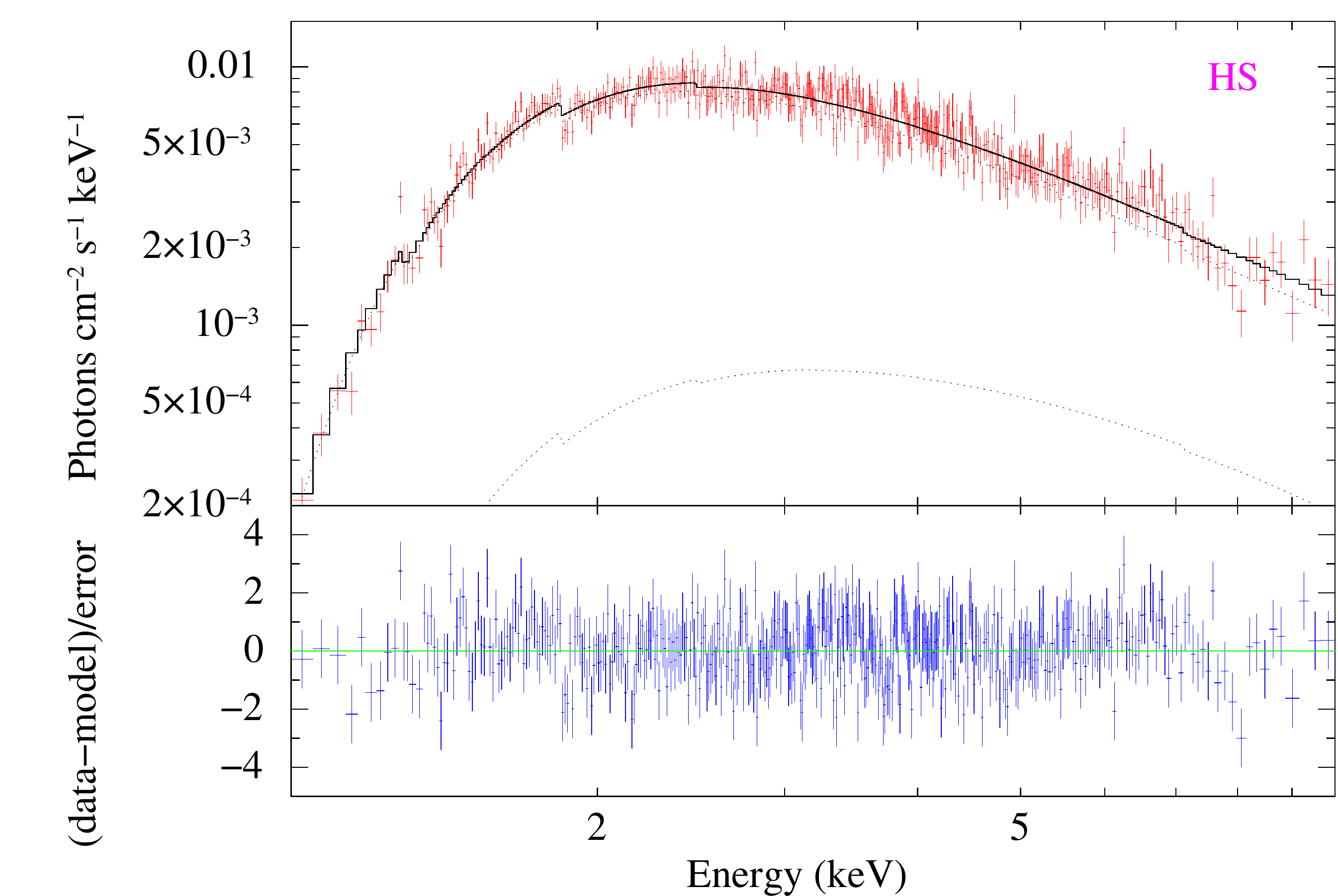}
\caption{Unfolded energy spectra of Swift J1728.9--3613 in the energy range 1--10 keV of {\it NICER} in different phases of the outburst. We have shown four spectra for the observation Ids 1200550101, 2200550106, 2200550131, and 2200550134, which correspond to different phases of the outburst, i.e., SIMS, SS, HIMS, and HS, respectively. Energy spectra are fitted with a combined model of {\it tbabs$\times$(nthcomp+diskbb)} and the residuals are shown in the bottom panels of each figure.}
\label{fig:spectrum}
}
\end{figure*}

\section{Discussion}
\label{sec:discussions}
In this paper, we study the evolution of different timing and spectral properties using multiple {\it NICER} and {\it Swift} observations of the newly discovered transient Swift J1728.9--3613 to identify the nature of the compact object. The outburst was characterized by a fast rise and a very slow decay of flux, which is typical for outbursts of X-ray binaries. During the study of timing evolution, we found a ``$q$'' shaped track in the HID traversing in the anticlockwise direction. This type of track is commonly observed during the outburst of black hole X-ray binaries \citep{Ho05, Re06}. In the RID, a partial hysteresis-like feature was also observed. This type of phenomenon in HID and RID are very much similar to be found in black hole X-ray binaries \citep{Be05, Re06, Mu11}. Although, $q$-shaped tracks are also observed in the HID of neutron star binaries \citep{Mu14, Ma22}, the combined study of timing and spectral properties along with the presence of type B QPOs in the PDS strongly suggest that the new transient Swift J1728.9--3613 is a black hole X-ray binary.

The source was discovered on MJD 58511.05 and at the time of its discovery, it was found to reach the near-peak flux observed by {\it Swift}/BAT (shown in Fig. \ref{fig:light}). From the study of combined timing and spectral properties, we found that the source underwent different spectral states at different phases of the outburst, which were similar to those observed in other black hole candidates.

We study the 1.5--9 keV XRT spectrum from the {\it Swift} observation on the day of its discovery. We found that the spectral index ($\Gamma=1.55\pm0.07$) and disc temperature were significantly low at the time of its discovery (see Table \ref{tab:parameters}). The PDS was dominated by a strong band-limited noise component. We did not detect any QPO from that observation. {\it MAXI}/GSC observed the field of the source before its discovery. Using the monitoring data of {\it MAXI}/GSC, it was evident from Fig. \ref{fig:light} that the flux started to increase from MJD 58508.5 and the corresponding HR was $\sim$1.4 which indicated that the source was in the HS when it was discovered. Thus, we concluded that the source remained in the HS from MJD 58508.50 to MJD 58511.05. A rapid decrease in HR was observed during that period.

On MJD 58512.64, the HR further decreased and the PDS was dominated by a broadband noise component with a typical fractional RMS of $\sim$8\%. Over the next few days, the fractional RMS varied between $\sim$1 and $\sim$4\%. Two QPOs were detected in the PDS at $\sim$5.40 Hz and $\sim$5.56 Hz respectively with decreasing RMS amplitude (see Table \ref{tab:qpo} for details). The shape of the PDS along with the presence of broadband noise component along with the frequencies, Q-values, and RMS amplitudes of the QPOs fitting parameters suggest that they were similar to the type-B QPOs which are generally found in the SIMS. At the same time, 1--10 keV energy spectra by {\it NICER} became steeper ($\Gamma=2.35\pm0.02$) and the disc temperature also increased (see Table \ref{tab:parameters}). The nature of the spectral shape and disc temperature associated with the presence of type-B QPOs in the PDS suggests that the source was in the SIMS on MJD 58512.64. The spectral index kept varying between $\sim$2.22 and $\sim$2.69 up to MJD 58533.17. Disk normalization kept on varying till MJD 58533.17, after which it became almost stable (see Fig. \ref{fig:parameters}) which is an indication that the source entered the SS after that day. The contribution of comptonized hard photon flux to the total flux was $\sim$90\% on MJD 58512.64, after which disk flux started dominating the spectra and the ratio $F_\text{nthcomp}/F_\text{total}$ was found to vary within 30--60 per cent. No evidence of HIMS was found during the rising phase of the outburst. The flux and disc temperature reached maxima on MJD 58521.05, followed by the gradual decay of both parameters that were shown in Fig. \ref{fig:parameters}. Radio observations were performed during that period with MeerKAT in multiple epochs \citep{Br19}. On the first MeerKAT observation on MJD 58514.11, the radio flux emitted from the source was 11.2$\pm$0.6 mJy whereas, on the latter two observations on MJD 58526.09 and MJD 58530.15, the source remained undetected \citep{Br19}. The source stayed in the SIMS for $\sim$22 days. The duration of SIMS is comparatively longer for this black hole Swift J1728.9--3613, which is rare. A similar long duration in SIMS was also observed in GX 339--4 during its 2002/2003 outburst \citep{Be05}.

The source was entered into the SS on MJD 58534.20. The spectral index increased to a higher value ($\Gamma=3.00\pm0.11$) and the corresponding HR decreased to $\sim$0.1 and it further decreased in the following few days. During that time, the fractional RMS dropped to a value as low as $\sim$1\%. A finite discontinuity in the model flux and disc temperature was observed in Fig. \ref{fig:parameters} during the transition. Spectra became steeper on the following days and kept varying within the range from $\sim$2.93 to $\sim$4.76 up to MJD 58612.13 indicating that the source remained in the same state up to that day. disc temperature and flux also decreased during that time. No evidence of type-A QPO was found in the PDS during this state. Disk photons were dominating the spectra, contributing $\sim$70\% of the total flux. Towards the end of the SS, $F_\text{nthcomp}/F_\text{total}$ started increasing again which suggests that comptonized hard photons became more dominant over the disk photons.

On the next {\it NICER} observation on MJD 58633.17, the spectral index decreased to $1.89\pm0.29$ and the corresponding fractional RMS became $\sim$14\% which further increased to $\sim$16\% on MJD 58636.01. The source seemed to return to the adjacent hard line (AHL) in the RID (Fig. \ref{fig:hrd_rid}) which was a clear indication that the source left the SS. Since no observation was performed between MJD 58612.13 and MJD 58633.17, the exact time of the transition is not found. The inner disk temperature became significantly low. No evidence of QPO was found between MJD 58633.17 and MJD 58636.01. The ratio $F_\text{nthcomp}/F_\text{total}$ was $\sim$60\% during the observation. The source was in the HIMS from MJD 58633.17 to MJD 58636.01.

The source was observed in the HS during the next {\it NICER} observation on MJD 58642.88 as the spectral index decreased to 1.88$\pm$0.05 and the fractional RMS further increased to $\sim$22\%. The HR increased to $\sim$0.8 which also indicates that the source reached the HS. The inner disc temperature became low as $\sim$0.276 keV and it further decreased on the following days (see Table \ref{tab:parameters}). 
The contribution of hard photon flux was 80--90\% of the total flux during that period. We did not find any evidence of SIMS in the decay phase. So, the source evolved through all four canonical states during the outburst of 2019 that are generally observed in black hole X-ray binaries \citep{Be05, Ho05, To20, Zh20}.

The {\it NICER} count rate suddenly increased on MJD 58576.01 which may be influenced by solar activity. Although we observed some certain changes in the spectral parameters during this event (see Tab. \ref{tab:parameters} and Fig. \ref{fig:parameters}). The spectral index decreased to a relatively lower value during that time and spectra became slightly harder. Inner disc temperature also increased slightly but the disk normalization did not vary much. It is possible that another outburst of a relatively smaller magnitude occurred which caused the change in other timing and spectral parameters during that period. A similar kind of re-flaring was observed during the decay phase of outbursts in several other sources, such as MAXI J1820+070 \citep{St20}, and MAXI J1348--630 \citep{Zh20, Sa21}.

In the SS, energy spectra were dominated by the soft component, which was associated with the blackbody radiation from an optically thick and geometrically thin accretion disc \citep{Sh73}. The inner disc radius ($r_{\text{in}}$) remained almost constant during that state and remained consistent with the innermost stable circular orbit (ISCO; \citet{Ta89, Eb93, Do07}), although the disc luminosity varied significantly. The inner disc radius (in the unit of km) can be calculated using the formula,
\begin{equation}
N_{\text{disc}} = \left( \frac{r_{\text{in}}}{D_{10}} \right)^2 \cos{i}
\end{equation}
where $N_{\text{disc}}$ is the {\it diskbb} normalization, $D_{10}$ is the source distance in the scale of 10 kpc and $i$ is the inclination of the disc. We estimated the corrected disc radius using the relation $R_{\text{in}}=\xi \kappa^2 r_{\text{in}}$, where $\xi=0.41$ is the correction factor \citep{Ku98}, and $\kappa=1.7$ is the colour hardening factor \citep{Sh95}. Same value of $\kappa$ was adopted to calculate the value of $R_{\text{in}}$ in some recent works in the literature \citep{To20, Zh20}. Here we considered the intermediate disc inclination of $i = 60$ degrees. From the fitting of the 1--10 keV {\it NICER} spectra in the SS, we obtained that the inner disc radius remained consistent within $35.72\pm0.58$ km and $45.26\pm2.27$ km, which is an indication of the standard accretion disc \citep{Sh73}. During the SS, the innermost disc radius corresponds to the ISCO. For a system of non-spinning black hole, the ISCO is equal to three times the Schwarzschild radius, i.e., $R_{\text{in}}=6GM_{\text{BH}}/c^{2}$. Using the median value of $R_{\text{in}}$ ($40.74\pm0.76$ km) obtained from the spectral fitting in SS and taking the source distance equal to 10 kpc, we calculated the approximate mass of the compact object to be $\sim$4.6 $M_\odot$. That was the lower limit of the black hole mass, considering the assumption of a non-spinning black hole. It would be more massive if the spin is considered. The inner disc temperature also remained as high as $\sim$1.058 keV when the luminosity of the source was relatively low (in the order of 10$^{37}$ erg s$^{-1}$, considering the distance equal to 10 kpc) at the beginning of the SS. High disc temperature and low luminosity indicate that Swift J1728.9--3613 might host a relatively low mass X-ray binary which is analogous to the estimated mass. Lower mass of Swift J1728.9--3613 categorized itself as one of the low mass X-ray binaries such as, GX 339--4 (2.3 to 9.5 $M_\odot$; \citet{He17}), 4U 1543--47 (2.7 to 7.5 $M_\odot$; \citet{Or98}).
Moreover, in the case of a black hole, soft to hard transition typically occurs at 1--4\% of the Eddington luminosity ($L_{\text{Edd}}$; \citet{Ma03}). We observed that in Swift J1728.9--3613, the soft-to-hard transition occurred when the luminosity was 0.01 $L_{\text{Edd}}$. This soft-to-hard transition occurred when the source flux was $\sim$4 per cent  of the peak flux. We found similar behavior in the case of other transients also. For example, in the case of MAXI J1348--630, the soft to hard transition occurred at 10 per cent  of the peak flux \citep{To20} and for MAXI J1820+070, it was about 12 per cent \citep{Sh19}.

\section{Conclusion}
\label{sec:conclusion}

We have studied different timing and spectral properties of the newly discovered X-ray transient Swift J1728.9--3613 using {\it NICER} and {\it Swift}/XRT observations.  The position of the source is measured at RA = $17^h28^m58.6401^s$, Dec = --36$^{\circ}$14${\arcmin}$35.321$\arcsec$ (J2000) using Chandra with enhanced accuracy. The HID shows a ``$q$''-shaped feature that traversed in the anticlockwise direction. 
A partial hysteresis is also observed in the RID. All these are typical characteristics of black hole transients. We have detected two type B QPOs in the SIMS of the outburst but did not find any evidence of type A and type C QPOs. The energy spectra are studied using XRT and {\it NICER}, from which we investigated the temporal evolution of spectral parameters. The evolution of the timing and spectral properties was similar to those found in other black hole X-ray binaries. In SS, the inner disc radius remained consistent with the ISCO which is a well-known characteristic of standard discs. We estimated the mass of the black hole to be 4.6 $M_\odot$ approximately, considering the source distance equal to 10 kpc without considering the spin of the system. The soft-to-hard transition occurred when the source luminosity was 0.01 $L_{\text Edd}$. The combined study of all these timings and spectral properties suggests that the new transient is a black hole X-ray binary.

\section*{Acknowledgements}
In this research, we used the {\tt HEASOFT v6.28} software for scientific data analysis, which was developed by the High Energy Astrophysics Science Archive Research Center (HEASARC) at NASA, GSFC. We used {\tt CIAO} developed by the {\it Chandra} X-ray Observatory team for the measurement of position. This research has made use of SAOImage {\tt DS9}, developed by the Smithsonian Astrophysical Observatory. The data products used in this work were collected from the HEASARC data archive. A special thanks to the {\it NICER} team, since their effort played a major role in this research. We would like to thank the Neil Gehrels {\it Swift} Observatory and MAXI-RIKEN teams since their daily monitoring data put a crucial impact on this work.

\section{Data Availability}
\label{sec:data}
All archival data used for the scientific analysis in this work are available in the HEASARC browse portal provided by High Energy Astrophysics Science Archive Research Center (HEASARC) at \href{https://heasarc.gsfc.nasa.gov/cgi-bin/W3Browse/w3browse.pl}{https://heasarc.gsfc.nasa.gov/cgi-bin/W3Browse/w3browse.pl}.



\begin{thebibliography}{99}

    \bibitem[{Arnaud}(1996)]{Ar96} Arnaud K. A., 1996, in Jacoby G. H., Barnes J., eds, ASP Conf. Ser. Vol. 101, Astronomical Data Analysis Software and Systems V. Astron. Soc. Pac., San Francisco, p. 17
    
    \bibitem[{Alam et al.}(2014)]{Al14} Alam M. S., Dewangan G. C., Belloni T., Mukherjee D., Jhingan S., 2014, MNRAS, 445, 4259A

   \bibitem[\protect\citeauthoryear{Bachetti et al.}{2015}]{Ba15} 
   Bachetti M., Harrison F.~A., Cook R., Tomsick J., Schmid C., Grefenstette B.~W., Barret D., et al., 2015, ApJ, 800, 109 
    \bibitem[{Barthelmy et al.}(2005)]{Ba05} Barthelmy S. D. et al., 2005, SSRv, 120, 143B

    \bibitem[{Barthelmy et al.}(2019)]{Ba19} Barthelmy S. D., D'Ai A., D'Elia V., Kennea J. A., Lien A. Y., Page K. L., Palmer D. M., Tohuvavohu A., 2019, The Astronomer's Telegram, 12436, 1
    
    \bibitem[{Belloni}(2009)]{Be09} Belloni T. M., 2009, Lecture Notes in Physics, 794, 53

    \bibitem [{Belloni et al.}(2005)]{Be05} Belloni T., Homan J., Casella P., van der Klis M., Nespoli E., Lewin W. H. G., Miller J. M., M\'{e}ndez M., 2005, A\&A, 440, 207

    \bibitem[{Belloni, Motta \& Mu\~{n}oz-Darias}(2011)]{Be11} Belloni T. M., Motta S. E., Mu\~{n}oz-Darias T., 2011, BASI, 39, 409B
    
    \bibitem[{Bright et al.}(2019)]{Br19} Bright J., Fender R., Woudt P., Miller-Jones J., 2019, The Astronomer's Telegram, 12522, 1
    
    \bibitem[{Burrows et al.}(2000)]{Bu00} Burrows D. N. et al., 2000, SPIE, 4140, 64B
    
    \bibitem[{Capitanio et al.}(2009)]{Ca09} Capitanio F., Belloni T., Del Santo M., Ubertini P., 2009, MNRAS, 398, 1194
    
    \bibitem[{Casella et al.}(2004)]{Ca04} Casella P., Belloni T., Homan J., Stella L., 2004, A\&A, 426, 587
    
    \bibitem[{Casella, Belloni \& Stella}(2005)]{Ca05} Casella P., Belloni T., Stella L., 2005, ApJ, 629, 403
    
    \bibitem[{Chakrabarti, Pal \& Nandi}(2006)]{Ch06} Chakrabarti S. K., Pal S., Nandi A., 2006, A\&A, 453, 965C
    
    \bibitem[{Del Santo et al.}(2016)]{De16} Del Santo M. et al., 2016, MNRAS, 456, 3585
    
    \bibitem[{Done, Gierli\'{n}ski \& Kubota}(2007)]{Do07} Done C., Gierli\'{n}ski M., Kubota A., 2007, A\&ARv, 15, 1
    
    \bibitem[{Dunn et al.}(2010)]{Du11} Dunn R. J. H. et al., 2010, MNRAS, 403, 61
    
    \bibitem[{Ebisawa et al.}(1993)]{Eb93} Ebisawa K. et al., 1993, ApJ, 403, 684
    
    \bibitem[{Enoto et al.}(2019)]{En19} Enoto T. et al., 2019, The Astronomer's Telegram, 12455, 1


    \bibitem[\protect\citeauthoryear{Garmire et al.}{2003}]{Ga03} 
    Garmire G.~P., Bautz M.~W., Ford P.~G., Nousek J.~A., Ricker G.~R., 2003, SPIE, 4851, 28 
    
    \bibitem[{Gendreau et al.}(2016)]{Ge16} Gendreau K. C. et al., 2016, in Proc. SPIE. p. 99051H, doi:10.1117/12.2231304
    
    \bibitem[{Gilfanov}(2010)]{Gi10} Gilfanov M., 2010, The Jet Paradigm from Microquasars to Quasars, Vol. 794, ed. T. Belloni (Berlin: Springer), 17
    
    \bibitem[{Heida et al.}(2017)]{He17} Heida M., Jonker P. G., Torres M. A. P., Chiavassa A., 2017, ApJ, 846, 132H
    
      \bibitem[{Homan \& Belloni}(2005)]{Ho05} Homan J., Belloni T., 2005, Ap\&SS, 300, 107
    
    \bibitem[{Homan et al.}(2001)]{Ho01} Homan J., Wijnands R., van der Klis M., Belloni T., van Paradijs J., Klein-Wolt M., Fender R., M\'{e}ndez M., 2001, ApJS, 132, 377
    
    \bibitem[{Hu et al.}(2019)]{Hu19} Hu Y. -D. et al., 2019, The Astronomer's Telegram, 12443, 1

    \bibitem[\protect\citeauthoryear{Ingram \& Done}{2012}]{In12} Ingram A., Done C., 2012, MNRAS, 419, 2369
    
    \bibitem[{Ingram \& Motta}(2019)]{In19} Ingram A. R. \& Motta S. E., 2019, NewAR, 8501524I
    
       
    \bibitem[{Kennea}(2019)]{Ke19} Kennea J. A., 2019, The Astronomer's Telegram, 12445, 1
    
    \bibitem[{Klein-Wolt \& van der Klis}(2008)]{Kl08} Klein-Wolt M., van der Klis M., 2008, ApJ, 675, 1407
    
    \bibitem[{Kubota et al.}(1998)]{Ku98} Kubota A. et al., 1998, PASJ, 50, 667
    
    \bibitem[{Leahy, Elsner \& Weisskopf}(1983)]{Le83} Leahy D. A., Elsner R. F., Weisskopf M. C., 1983, ApJ, 272, 256L
    
    \bibitem[{Maccarone}(2003)]{Ma03} Maccarone T. J., 2003, A\&A, 409, 697
    
    \bibitem[{Makishima et al.}(1986)]{Ma86} Makishima K., Maejima Y., Mitsuda K., Bradt H. V., Remillard R. A., Tuohy I. R., Hoshi R., Nakagawa M., 1986, ApJ, 308, 635M
    
    \bibitem[{Mandal \& Pal}(2022)]{Ma22} Mandal M., Pal S., 2022, MNRAS, 511, 1121
    
    \bibitem[{Mitsuda et al.}(1984)]{Mi84} Mitsuda K. et al., 1984, PASJ, 36, 741M
    
    \bibitem[{Motta}(2016)]{Mo16} Motta S. E., 2016, Astronomische Nachrichten, 337, 398
    
    \bibitem[{Motta, Belloni \& Homan}(2009)]{Mo09} Motta S., Belloni T., Homan J., 2009, MNRAS, 400, 1603
    
    \bibitem[{Motta et al.}(2011)]{Mo11} Motta S., Mu\~{n}oz-Darias T., Casella P., Belloni T., Homan J., 2011, MNRAS, 418, 2292
    
    \bibitem[{Mu\~{n}oz-Darias, Motta \& Belloni}(2011)]{Mu11} Mu\~{n}oz-Darias T., Motta S., Belloni T. M., 2011, MNRAS, 410, 679
    
    \bibitem[{Mu\~{n}oz-Darias et al.}(2014)]{Mu14} Mu\~{n}oz-Darias T., Fender R. P., Motta S. E., Belloni T. M., 2014, MNRAS, 443, 3270M
    
    \bibitem[{Negoro et al.}(2019)]{Ne19} Negoro H. et al., 2019, The Astronomer's Telegram, 12437, 1
    
    \bibitem[{Orosz et al.}(1998)]{Or98} Orosz J. A., Jain R. K., Bailyn C. D., McClintock J. E., Remillard R. A., 1998, ApJ, 499, 375O
    
    \bibitem[{Pal et al.}(2006)]{Pa06} Pal S., Chakrabarti S. K., Kraus A., Mandal S., 2006, BASI, 34, 1P
    
    \bibitem[{Poole et al.}(2008)]{Po08} Poole T. S. et al., 2008, MNRAS, 383, 627P 

    \bibitem[\protect\citeauthoryear{Remillard et al.}{2021}]{Re21}
Remillard R.~A. et al., 2021, preprint (arXiv:2105.09901)

	\bibitem[{Remillard \& McClintock}(2006)]{Re06} Remillard R. A., McClintock J. E., 2006, ARA\&A, 44, 49
	
    \bibitem[{Saha et al.}(2021)]{Sa21} Saha D., Pal S., Mandal M., Manna A., 2021, arXiv e-prints, p. arXiv:2104.09926
	
	\bibitem[{Shakura \& Sunyaev}(1973)]{Sh73} Shakura N. I., Sunyaev R. A., 1973, A\&A, 500, 33
	
	\bibitem[{Shidatsu et al.}(2019)]{Sh19} Shidatsu M. et al., 2019, ApJ, 874, 183
	
	\bibitem[{Shimura \& Takahara}(1995)]{Sh95} Shimura T., Takahara F., 1995, ApJ, 445, 780
	
	\bibitem[{Skrutskie et al.}(2006)]{Sk06} Skrutskie M. F. et al., 2006, AJ, 131, 1163S 
	
	\bibitem[{Sriram, Rao \& Choi}(2013)]{Sr13} Sriram K., Rao A. R., Choi C. S., 2013, ApJ, 775, 28
	
	\bibitem[{Stevans \& Uttley}(2016)]{St16} Stevens A. L., Uttley P., 2016, MNRAS, 460, 2796
	
	\bibitem[{Stiele \& Kong}(2020)]{St20} Stiele H., Kong A. K. H., 2020, ApJ, 889, 142
	
	\bibitem[{Sturner \& Shrader}(2005)]{St05} Sturner S. J., Shrader C. R., 2005, ApJ, 625, 923
	
	\bibitem[{Tanaka}(1989)]{Ta89} Tanaka Y., 1989, in ESA Special Publication, Vol. 1, Two Topics in X-Ray Astronomy, Volume 1: X-Ray Binaries, Volume 2: AGN and the X-Ray Background, ed. J. Hunt \& B. Battrick, 3
	
	\bibitem[{Tominaga et al.}(2020)]{To20} Tominaga M. et al., 2020, ApJ, 899, 20
	
	\bibitem[{van der Klis}(2006)]{Va06} van der Klis M., 2006, Rapid X-ray Variability. pp 39–112
	
	\bibitem[{Wijnands, Homan \& van der Klis}(1999)]{Wi99} Wijnands R., Homan J., van der Klis M., 1999, ApJ, 526, L33
	
	\bibitem[{Wilms, Allen \& McCray}(2000)]{Wi00} Wilms J., Allen A., McCray R., 2000, ApJ, 542, 914
	
	\bibitem[{Zdziarski, Johnson \& Magdziarz}(1996)]{Zd96} Zdziarski A. A., Johnson W. N., Magdziarz P., 1996, MNRAS, 283, 193Z
	
	\bibitem[{Zhang et al.}(2020)]{Zh20} Zhang L. et al., 2020, MNRAS, 499, 851Z
	
	\bibitem[{\.{Z}ycki, Done \& Smith}(1999)]{Zy99} \.{Z}ycki P. T., Done C., Smith D. A.,  1999, MNRAS, 309, 561Z

\end{thebibliography}
\end{document}